\documentclass[english,aps,prl,amsmath,twocolumn,reprint,superscriptaddress,notitlepage]{revtex4-2}
\usepackage[T1]{fontenc}

\makeatletter

\usepackage{graphicx}
\graphicspath{ {figures/}{figures/eps/}{figures/pdf/} }
\usepackage{dcolumn}
\usepackage{graphicx}
\usepackage{xcolor} 

\usepackage{hyperref}
\hypersetup{
 linkcolor=blue,urlcolor=blue,citecolor=blue}

\usepackage{nomencl}                    
\usepackage[colorinlistoftodos]{todonotes}
\usepackage{subcaption}

\usepackage{fancyhdr}                   
\usepackage{url}                        
\usepackage[inactive]{srcltx}		 	
\usepackage{relsize}                    
\usepackage{booktabs}                   
\usepackage{mathrsfs}                   
\usepackage[titletoc]{appendix}			
\usepackage{comment}
\usepackage{dsfont}
\usepackage{amsmath}
\usepackage{mathtools}
\usepackage{newtxmath}

\usepackage{amssymb}
\usepackage{array}
\usepackage[utf8]{inputenc}
\usepackage{color}
\usepackage{xcolor}
\usepackage{bm}
\usepackage{subcaption}
\usepackage{wrapfig}
\usepackage{braket}

\usepackage{feynmp-auto}
\usepackage[compat=1.1.0]{tikz-feynman}
\tikzfeynmanset{warn luatex=false} 
\usepackage{siunitx}
\usepackage{ulem}
\usepackage{qcircuit}
\DeclarePairedDelimiterXPP\BigO[1]{\mathcal{O}}{\lparen}{\rparen}{}{#1}
\DeclareUnicodeCharacter{2212}{-}
\DeclareUnicodeCharacter{2265}{\ensuremath{\geq}}
\usepackage{tikz}
\usetikzlibrary{shapes.geometric, arrows.meta, positioning, calc, matrix}

\begin{document}

\title{Quantum Co-Design of Inhomogeneous Many-Body Neutrino Fast Flavor Transformation}
\author{Zoha Laraib}
\email{zlaraib@vols.utk.edu}
\affiliation{Department of Physics and Astronomy, The University of Tennessee, Knoxville, Tennessee 37996, USA}
\affiliation{National Centre for Computer Science, Oak Ridge National Laboratory, Oak Ridge, Tennessee 37830, USA}
\author{Sherwood Richers}
\affiliation{Department of Physics and Astronomy, The University of Tennessee, Knoxville, Tennessee 37996, USA}
\author{Alessandro Baroni}
\affiliation{National Centre for Computer Science, Oak Ridge National Laboratory, Oak Ridge, Tennessee 37830, USA}
\author{Kathleen Hamilton}
\affiliation{Computer Science and Engineering Division, Oak Ridge National Laboratory, Oak Ridge, Tennessee 37830, USA}
\author{In-Saeng Suh}
\affiliation{National Centre for Computer Science, Oak Ridge National Laboratory, Oak Ridge, Tennessee 37830, USA}

\date{\today}

\begin{abstract}
Dense-neutrino flavor evolution is a quantum many-body problem whose fully
correlated treatment becomes rapidly more expensive with increasing particle
number and spatial structure. We
develop a \texttt{QCNO} quantum simulation code that maps a
inhomogeneous forward-scattering neutrino Hamiltonian with advection and
finite-range interactions to TEBD2 product-formula quantum circuits, and connects ideal
many-body simulation, backend-aware execution, and fault-tolerant resource
estimation within the same physical model. We reproduce the exact many-body evolution of the suppressed mean-field-like
transverse fast flavor instability through $N=30$ and show that even a single
open-boundary interaction generates R\"enyi entanglement and
non-stabilizer magic, although noisy backends still produce polarization RMSEs of order
$0.1$--$0.3$. The restricted active interaction graph of this problem allows us to estimate the circuit depth and $T$ gate cost through $N=50$ for both NISQ and fault-tolerant approaches. The measured TEBD2 error in our fiducial ideal simulation of order $10^{-3}$ motivates a
synthesis tolerance $\varepsilon_{\rm syn}\sim10^{-7}$, corresponding to about
$70$ $T$ gates per $R_Z$ rotation. The generated circuit contains $1.4\times10^4$ $T$ gates per qubit for open boundaries (twice that for closed boundaries), requiring an application-level logical-$T$ error
target of order $10^{-7}$ for early fault-tolerant architectures.
\end{abstract}

\maketitle


A central challenge in quantum simulation is to determine how classical and
quantum resources should be allocated across structured classical algorithms,
noisy quantum processors, and fault-tolerant logical hardware for a specified
physical observable and target accuracy~\cite{feynman1982simulating}. Symmetries, sparsity, locality, and
limited entanglement can keep many-body (MB) problems accessible to specialized
classical methods~\cite{vidal2003efficient,schollwock2011density}, whereas dynamics that defeat such compression
require deeper quantum circuits and are consequently more exposed to noise~\cite{preskill2018quantum}.
Useful quantum simulation therefore requires application-level co-design, in
which the physical model, Hamiltonian-simulation algorithm, measurement
strategy, compiler, and hardware architecture are optimized jointly around
the scientifically relevant observables and a controlled error and resource
budget~\cite{bauer2023hep,babbush2018encoding}. This co-design must also connect near-term noisy
execution to the logical resources required for quantitatively controlled
fault-tolerant simulation~\cite{campbell2017roads}.

Spatially inhomogeneous neutrino flavor evolution provides a challenging 
application through which to study this transition. Core-collapse supernovae can emit of order \(10^{58}\) neutrinos and
antineutrinos, while neutron-star mergers likewise produce dense neutrino
fields whose transport of energy and lepton number influences ejecta
composition, nucleosynthesis, and observable signals~\cite{janka_ExplosionMechanismsCoreCollapse_2012,mezzacappa_CoreCollapseSupernova_2023,foucart2023neutrino,burrows2021core,Qian1993,wang2022neutrinos,fischer2024neutrinos, balantekin2024collective}. Coherent neutrino-neutrino forward scattering couples the flavor degrees of freedom of different particles and generates an
interacting quantum MB system whose entanglement closure produces mean-field (MF) approximation of
nonlinear collective evolution~\cite{pantaleone1992neutrino,vlasenko2014neutrino, duan2010collective,tamborra2021new,capozzi2022neutrino,richers_FastFlavorTransformations_2022,volpe2024neutrinos,johns2025neutrino,patwardhan_ManybodyCollectiveNeutrino_2023}. In this limit, of particular interest are fast flavor instabilities, whose growth rates
are set by the neutrino self-interaction scale and which may modify the
neutrino radiation field and ejecta composition~\cite{sawyer2005speed,sawyer2009multiangle,banerjee2011linearized,dasgupta2017fast,wu2017imprints,li2021neutrino,just2022fast,fernandez_FastFlavorInstability_2022a,ehring2023fast,wang2025effect}. Realistic environments are spatially inhomogeneous, so neutrinos cannot generally be treated as identically coupled particles in a
fully collective model. These ingredients
determine both the strength and connectivity of the many-body interaction and
can qualitatively modify the MF instability~\cite{
sigl1993general,mckellar1994oscillating,Shalgar2021breaking,fiorillo2024fast,fiorillo2024inhomogeneous}.

Previous studies have made the many-body neutrino problem tractable by
introducing complementary approximations on classical and quantum hardware. On classical hardware, tensor-network methods---particularly
matrix product states (MPSs) evolved with time-evolving block decimation and related
algorithms~\cite{orus2014practical, vidal2004efficient,vidal2007classical,verstraete2008matrix}---have extended simplified collective and few-beam calculations
to as many as around thousand neutrinos and provided benchmarks for
entanglement growth, equilibration, and departures from MF
dynamics~\cite{roggero2021entanglement,roggero2021dynamical,roggero2022entanglement,martin2022classical,Cervia_2022,Siwach_2023}. This reach relies on representing the evolving state at a limited bond dimension to approximate it
by compressing the exact \(2^N\) amplitudes that constitute the wave function.
Position-dependent kernels, finite overlap functions, angular couplings, advection, boundary conditions, and particle reordering produce a time-dependent interaction graph in which strongly coupled particles are not necessarily adjacent in the one-dimensional tensor-network ordering. This destroys the effective locality that allows an MPS to represent the evolving state with a modest bond dimension, causing entanglement to span many tensor-network cuts and making controlled compression substantially more difficult. Recent inhomogeneous tensor-network calculations that used the truncated Hamiltonian to incorporate these effects did not clearly recover the coherent fast instability predicted by MF theory within the accessible particle numbers and bond dimensions~\cite{laraib2025many}.
Meanwhile, neutrino studies on quantum hardware have demonstrated the execution of systems with tens of particles for homogeneous, fully collective, or symmetry-reduced Hamiltonians~\cite{hall2021simulation,spagnoli2025collective,yeteraydeniz2021collective,siwach2023collective,turro2025qutrit, heimsoth2026three, martin2022classical,xiong2022many,illa2023multi}. Thus, classical methods reach larger \(N\) under compressible interaction
geometries, while quantum hardware natively represents smaller MB
systems under substantially simplified conditions. The spatially
inhomogeneous regime remains unexplored on quantum hardware, leaving
an open question of whether the suppression of FFI-like growth is intrinsic to the
MB dynamics or reflects finite bond
dimension, entanglement, and favorable particle ordering.

The two-flavor problem nevertheless has a natural digital encoding: each
neutrino maps to one qubit, vacuum evolution to one-qubit rotations, and
neutrino-neutrino scattering to two-qubit spin interactions, so an
\(N\)-particle system maps directly to an \(N\)-qubit register~\cite{pehlivan2011invariants,hall2021simulation}. This avoids the explicit classical storage or compression of the full
MB wave function, but does not guarantee efficient calculations.  Consequently, the spatially dependent interaction kernel and geometry control the active
two-qubit gate count, routing overhead, circuit depth, sensitivity to
backend noise, and logical resource cost~\cite{cowtan2019qubit,preskill2018quantum}, while the measurement protocol and
shot count determine the statistical precision~\cite{huggins2021efficient}.

Due to the deep circuits, this application spans the transition from noisy
intermediate-scale quantum (NISQ) execution to fault-tolerant computation~\cite{preskill2018quantum,campbell2017roads}. At
small system sizes, noisy execution can validate the encoding, benchmark native-gate
decompositions, quantify routing and measurement overheads, and determine
whether the physical signal survives realistic noise. Scaling to larger
systems, longer evolution times, and controlled accuracy requires
error-corrected logical execution, where arbitrary rotations must be
synthesized in a discrete universal gate set and non-Clifford operations, become a principal resource~\cite{bravyi2005universal,selinger2015efficient,
campbell2017roads}.
We use direct
product-formula Hamiltonian simulation as a common algorithmic layer across
these regimes and evolve the
truncated Hamiltonian through Trotter time-slicing
\(e^{-iHt}\)~\cite{lloyd1996universal}, instead of a variational ansatz or outer
classical optimization loop.
The same physics-defined circuit can therefore
be executed on noisy hardware and decomposed into logical Clifford+\(T\)
resources. More advanced methods based on block encoding, qubitization, or
quantum signal processing may improve asymptotic precision scaling, but
require additional logical primitives whose potential must be evaluated
within the complete application-level resource model~\cite{low2017optimal,low2019hamiltonian,gilyen2019quantum,clinton2021hamiltonian,bauer2023hep}.

In this work, we develop \texttt{QCNO} (\textit{Quantum Computing for
Neutrino Oscillations}), a modular framework for
simulation of spatially inhomogeneous MB neutrino flavor evolution using a truncated Hamiltonian.
We use \texttt{QCNO} to connect flavor dynamics with application, hardware,
and fault-tolerant co-design. We compare finite-system MB evolution
with MF fast-flavor dynamics, characterize the generated hierarchy of
entanglement and stabilizer magic, determine how interaction geometry and
backend architecture control gate counts and circuit resources, and translate
the same Trotterized circuits into accuracy-dependent Clifford+\(T\)
estimates. Product-formula convergence, finite-shot uncertainty,
transpilation, connectivity, and backend noise are incorporated into the
simulation error and resource model. The resulting workflow characterizes where spatially inhomogeneous neutrino
simulation encounters the limitations of classical Tensor network compression and noisy
hardware and establishes the error and logical-resource requirements for
extending it toward fault-tolerant execution.

The paper is organized as follows. In
Sec.~\ref{sec:quantum_framework}, we define the \texttt{QCNO} framework for the inhomogeneous Hamiltonian and
its qubit encoding and describe the product-formula evolution, advection, and
backend execution. In
Sec.~\ref{sec:results}, we examine the flavor dynamics and finite-size
behavior, analyze entanglement and stabilizer-R\'enyi structure, quantify
backend-native and fault-tolerant resources, and assess numerical and
hardware robustness. We conclude by discussing the application, hardware,
and fault-tolerant co-design requirements for extending spatially
inhomogeneous MB simulations beyond the reliable reach of current
classical and noisy quantum methods.

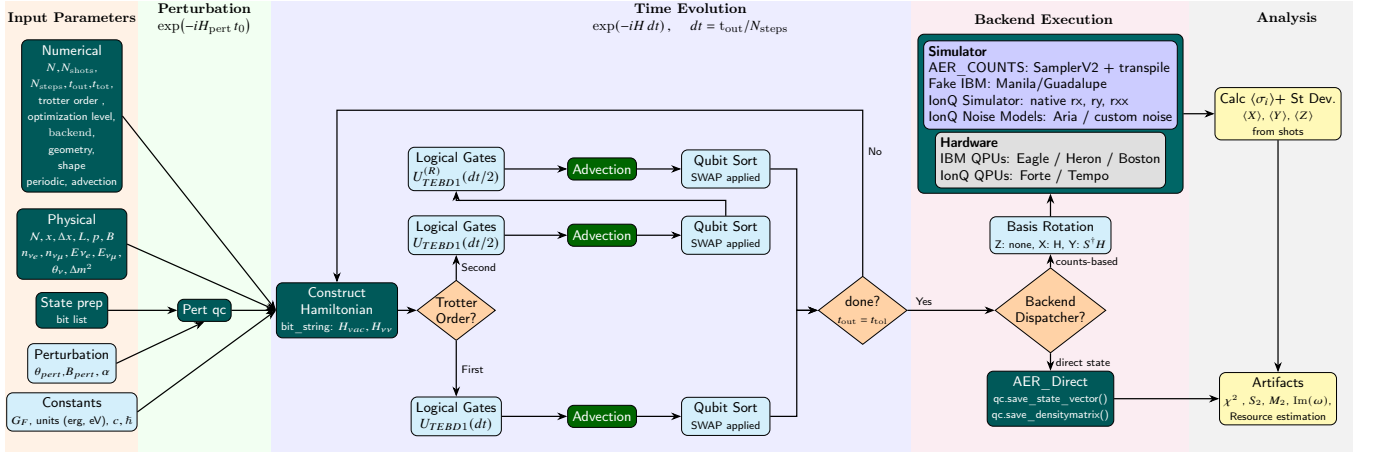
\begin{figure*}[t!]
\centering
\resizebox{\textwidth}{!}{%
\begin{tikzpicture}[
    font=\sffamily\small,
    >=Stealth,
    block/.style={rectangle, rounded corners, draw=black, fill=cyan!15,
        minimum width=2cm, minimum height=0.5cm, align=center},
    main/.style={rectangle, rounded corners, draw=black, fill=teal!70!black,
        text=white, minimum width=1.0cm, minimum height=0.5cm, align=center},
    decision/.style={diamond, draw=black, fill=orange!35, aspect=1.3,
        minimum width=0.5cm, inner sep=0.3pt, align=center},
    backend/.style={rectangle, rounded corners, draw=black, fill=pink!45,
        minimum width=1.5cm, minimum height=0.3cm, align=center},
    simulator/.style={backend, fill=blue!20},
    hardwarebox/.style={backend, fill=orange!25},
    output/.style={rectangle, rounded corners, draw=black, fill=yellow!35,
        minimum width=2.0cm, minimum height=0.5cm, align=center},
    arrow/.style={->, thick}
]
\draw[fill=orange!10, draw=none] (-0.5,4.5) rectangle (2.5,-5.7);
\draw[fill=green!5, draw=none] (2.5,4.5) rectangle (5.5,-5.7);
\draw[fill=blue!7, draw=none] (5.5,4.5) rectangle (20,-5.7);
\draw[fill=purple!7, draw=none] (20,4.5) rectangle (26.3,-5.7);
\draw[fill=gray!10, draw=none] (26.3,4.5) rectangle (30,-5.7);
\node[font=\bfseries, align=center] at (1.0,4.1) {Input Parameters};
\node[font=\bfseries, align=center] at (4.0,4.1) {Perturbation\\ $\exp\!\left(-i H_{\mathrm{pert}}\, t_0\right)$};
\node[font=\bfseries, align=center] at (15,4.1) {Time Evolution\\ $\exp\!\left(-i H\,dt\right),\quad dt=\mathrm{t_{out}}/N_{\mathrm{steps}}$};
\node[font=\bfseries, align=center] at (23.0,4.1) {Backend Execution};
\node[font=\bfseries, align=center] at (28.5,4.1) {Analysis};
\node[main]  (numerical) at (1.0,1.9) {Numerical\\ \scriptsize $N,$$N_\mathrm{shots}$,\\ \scriptsize$  N_\mathrm{steps},t_\mathrm{out},$$t_\mathrm{tot}$,\\ \scriptsize  trotter order ,\\ \scriptsize optimization level,   \\ \scriptsize $\mathrm{backend}$, \\ \scriptsize geometry,\\ \scriptsize shape\\ \scriptsize  periodic, advection};
\node[main]  (physical) at (1.0,-1.0) {Physical\\ \scriptsize $\mathcal{N},x,\Delta x,L,p,B$\\ \scriptsize $n_{\nu_e},n_{\nu_\mu}, E{\nu_e}, E_{\nu_\mu}, $\\ \scriptsize $ \theta_{\nu}, \Delta{m^2}$};
\node[main] (state_prep) at (1.0,-2.5) {State prep\\ \scriptsize {bit list}};
\node[block] (pert_consts)  at (1.0,-3.7) {Perturbation\\ \scriptsize $\theta_{pert}$,$B_{pert},  \alpha$};
\node[block] (constants)  at (1.0,-4.8) {Constants \\ \scriptsize $G_F$, units (erg, eV), $c, \hslash$};
\node[main]  (pert) at (4.0,-2.5) {Pert qc};
\node[main] (ham)    at (7.0,-2.5) {Construct\\ Hamiltonian\\ \scriptsize bit\_string: $H_{vac}, H_{\nu \nu}$};
\node[decision] (trotter) at (9.7,-2.5) {Trotter\\Order?};
\node[block] (first) at (9.7,-4.9) {Logical Gates\\ $U_{TEBD1}(dt)$};
\node[block] (low1) at (9.7,-0.8) {Logical Gates\\ $U_{TEBD1}(dt/2)$};
\node[main, fill=green!40!black] (low2) at (13,-0.8) {Advection};
\node[block] (low3) at (9.7,0.7) {Logical Gates\\ $U_{TEBD1}^{(R)}(dt/2)$};
\node[main, fill=green!40!black] (low4) at (13,0.7) {Advection};
\node[main, fill=green!40!black] (adv) at (low2 |- first) {Advection};
\node[block] (sortB) at (15.8,0.7) {Qubit Sort\\ \scriptsize SWAP applied};
\node[block] (sortC) at (15.8,-4.9) {Qubit Sort\\ \scriptsize SWAP applied};
\node[block] (sortA) at (15.8,-0.8) {Qubit Sort\\ \scriptsize SWAP applied};
\node[decision] (confirm) at (18.9,-2.5) {done? \\ \scriptsize $t_\mathrm{out}= t_{\rm tol}$};
\node[decision] (dispatch) at (23.15,-2.5) {Backend\\Dispatcher?};
\node[block] (basisrot) at (23.15,-0.8) {Basis Rotation\\ \scriptsize Z: none,\ X: H,\ Y: $S^\dagger H$};
\node[main] (directstate) at (23.15,-4.5) {AER\_Direct\\ \scriptsize qc.save\_state\_vector()\\ \scriptsize qc.save\_density\-matrix()};
\node[draw=black, rounded corners, fill=teal!120, minimum width=6cm, minimum height=3.6cm] (execution) at (23.15,1.95) {};
\node[draw, rounded corners, fill=blue!20, minimum width=1.2cm, align=left] (execution_sim) at (23.15,2.6)
{\textbf{Simulator}\\ AER\_COUNTS: SamplerV2 + transpile\\ Fake IBM: Manila/Guadalupe\\ IonQ Simulator: native rx, ry, rxx\\ IonQ Noise Models: Aria / custom noise};
\node[draw, rounded corners, fill=gray!25, minimum width=1.2cm, align=left] (execution_hw) at (23.150,0.9)
{\textbf{Hardware}\\ IBM QPUs: Eagle / Heron / Boston \\ IonQ QPUs: Forte / Tempo};
\node[output] (counts) at (28.3,1.95) {Calc $\langle\sigma_i\rangle$+ St Dev.\\ \scriptsize $\langle X \rangle$, $\langle Y \rangle$, $\langle Z \rangle$ \\ \scriptsize from shots};
\node[output] (stats) at (28.3,-4.5) {Artifacts \\ \scriptsize $\chi^2$ , $S_2$, $M_2 $, $\mathrm{Im}(\omega)$,\\ \scriptsize Resource estimation};
\draw[arrow] (numerical.east) -- (ham.west);
\draw[arrow] (physical.east) -- (ham.west);
\draw[arrow] (constants.east) -- (ham.west);
\draw[arrow] (pert_consts.east) -- (pert.south);
\draw[arrow] (state_prep) -- (pert);
\draw[arrow] (pert) -- (ham);
\draw[arrow] (ham.east) -- ++(0,0) |- (trotter.west);
\draw[arrow] (trotter) -- node[right] {\scriptsize First} (first);
\draw[arrow] (trotter) -- node[right] {\scriptsize Second} (low1);
\draw[arrow] (low1) -- (low2);
\draw[arrow] (low2) -- (sortA);
\draw[arrow] (sortA.north) -- ++(0,0.35) -| (low3.south);
\draw[arrow] (low3) -- (low4);
\draw[arrow] (low4) -- (sortB);
\draw[arrow] (sortB.east) -- ++(0.6,0) |- (confirm.west);
\draw[arrow] (first) -- (adv);
\draw[arrow] (adv) -- (sortC);
\draw[arrow] (sortC.east) -- ++(0.6,0) |- (confirm.west);
\draw[arrow] (confirm.east) -- node[above,pos=0.75] {\scriptsize Yes} ++(0.5,0) |- (dispatch.west);
\draw[arrow] (confirm.north) -- node[right,pos=0.75] {\scriptsize No} ++(0,3.75) -| (ham.north);
\draw[arrow] (dispatch.north) -- node[right,pos=0.5] {\scriptsize counts-based} (basisrot.south);
\draw[arrow] (dispatch.south) -- node[right,pos=0.5] {\scriptsize direct state} (directstate.north);
\draw[arrow] (basisrot.north) -- (execution.south);
\draw[arrow] (execution.east) -- (counts.west);
\draw[arrow] (counts) -- (stats);
\draw[arrow] (directstate) -- (stats);
\end{tikzpicture}

}
\caption{QCNO code framework workflow showing the full pipeline for a test. From parameter initialization to applying perturbation, time evolution via Hamiltonian construction and gate application, backend execution across simulators and quantum hardware, measurement processing, statistical analysis, and PDF artifact archiving. Particle sorting based on positions is applied in every layer (showed by colored background).}
\label{fig:qcno_code_framework}
\end{figure*}

\section{QCNO Framework}
\label{sec:quantum_framework}
The \texttt{QCNO} framework begins from the physical neutrino Hamiltonian, maps the spin--flavor degrees of freedom to qubits, decomposes the Hamiltonian into a list of one-site and two-site terms, constructs Trotterized quantum circuits, optionally includes perturbations and spatial advection, dispatches the resulting circuits to backend-specific execution paths, and finally reconstructs physical observables and quantum-resource diagnostics. The full computational workflow is shown schematically in Fig.~\ref{fig:qcno_code_framework}.

\paragraph{Physical model.}

We evolve a MB neutrino system composed of many computational particles, each of which represents a finite volume of phase space. Within the two-flavor approximation, each particle is mapped to a single qubit,
\begin{equation}
    |\nu\rangle
    =
    \alpha|\nu_e\rangle+\beta|\nu_x\rangle,
    \qquad
    |\nu_e\rangle\equiv |0\rangle,
    \qquad
    |\nu_x\rangle\equiv |1\rangle .
\end{equation}
A general one-qubit flavor state is therefore written as
\begin{equation}
    |\psi\rangle
    =
    \cos\frac{\theta}{2}|0\rangle
    +
    \sin\frac{\theta}{2}e^{i\phi}|1\rangle ,
\end{equation}
with the Bloch-vector components corresponding to the local flavor-polarization components. The circuit is always prepared in a computational-basis product state, so the general one-site state above describes the evolved state rather than the initial condition. 

Particles are sorted on qubits before each time step according to their physical position. The IO routines track a particle's qubit index over time so output quantities are always ordered according to the original site order rather than their current order.

The time-evolution of the state is governed by the Scr\"odinger equation including contributions to the Hamiltonian from vacuum oscillations and neutrino-neutrino interactions,
\begin{equation}
    H = H_{\mathrm{vac}} + H_{\nu\nu} .
    \label{eq:qcno_full_hamiltonian}
\end{equation}
The vacuum Hamiltonian is implemented as a one-body flavor-precession term~\cite{patwardhan_ManybodyCollectiveNeutrino_2023},
\begin{equation}
    H_{\mathrm{vac}}
    = \sum_i h_i =
    \frac{1}{2} \sum_i
    s_i\omega_i\,\vec B\cdot\vec \sigma_i ,
    \label{eq:qcno_vacuum}
\end{equation}
where $i\in\{1,\dots,N\}$ indicates which spin-1/2 degree of freedom is being operated on. The vacuum oscillation frequency
$
    \omega_i=\frac{\Delta m^2}{2|E_i|}
$
and the vacuum mixing vector
$
    \vec B=(\sin2\theta_\nu,0,-\cos2\theta_\nu)
$
depend on the mass-squared splitting $\Delta m^2$, the mixing angle $\theta_\nu$, and the particle momenta $\vec{p}_i$ (where the energy is $E_i=|\vec{p}_i|$). We also implement an energy sign $s_i\in\{-1,1\}$ for each neutrino to allow us to replicate results in the literature, though $s_i=1$ in all calculations in the main text. The spin (i.e., flavor) vector of operators acting on a site $i$ is $\vec\sigma_i=(\sigma_i^x,\sigma_i^y,\sigma_i^z)$.

The MB neutrino-neutrino (truncated forward-scattering) Hamiltonian is written as\cite{laraib2025two,patwardhan_ManybodyCollectiveNeutrino_2023}
\begin{equation}
    H_{\mathrm{\nu\nu}}
    =\sum_{i<j} h_{ij}=
    \frac{\mu}{2}
    \sum_{i<j}
    \vec\sigma_i\cdot\vec\sigma_j\,
    J_{ij}\,
    S(\xi_{ij}).
    \label{eq:qcno_self_interaction}
\end{equation}
The interaction scale $
    \mu=\sqrt{2}G_F n
$ is an input parameter choice for a given neutrino number density $n$.
Spatial inhomogeneity is included through the shape function $S(\xi_{ij})$, where $\xi_{ij}=(x_i-x_j)/\Delta_x$ is the normalized distance between two particles. The homogeneous case corresponds to $
    S(\xi)=1$ (i.e., each computational particle has infinite extent),
while compact-support kernels, such as the triangular kernel,
\begin{equation}
    S(\xi)
    =
    \left(1-|\xi|\right)
    \Theta(1-|\xi|),
\end{equation}
encode finite computational particle size and spatial separation. In the homogeneous isotropic limit, Eq.~\eqref{eq:qcno_self_interaction} reduces to the standard all-to-all collective-neutrino Hamiltonian. Particles can be positioned at any position $x_i$; the parameter $\Delta_x$ just governs the range of interactions. When using periodic boundary conditions, if $|x_i-x_j|>L/2$ we use $\xi_{ij} \leftarrow \xi_{ij} \pm L/\Delta_x$ to cause the shape function to wrap around the periodic boundary. The angular factor $
    J_{ij}=1-\hat p_i\cdot\hat p_j
$
captures the angular dependence of the weak interaction. The isotropic limit is obtained by setting $J_{ij}=1$. 

\paragraph{Input initialization.}

The input layer of Fig.~\ref{fig:qcno_code_framework} initializes both physics and execution parameters. The numerical parameters include the number of sites $N$, the measurement interval $\mathrm{t_{out}}$, the number of Trotter substeps in each measurement interval $\mathrm{N_{steps}}$, the Trotter order(trotter order), the total evolution time $\mathrm{t_{tot}}$, the backend name, the transpiler optimization level, the number of shots, the geometry selector, the shape selector, and the booleans controlling periodic boundaries and advection.

Particles are initialized using a list of positions $\vec{x}_i$, a list of momenta $\vec{p}_i$, and a list of initial flavor states, from which a MB product state is constructed. 

\paragraph{Perturbation.}

Before the main Hamiltonian evolution, QCNO can apply a controlled perturbation, 
\begin{equation}
    U_{\mathrm{pert},i}
    =
    \prod_{d\in\{x,y,z\}}e^{-i\alpha B^d_\mathrm{pert}\sigma^d_i/2}
\end{equation}
where $\alpha$ controls the size of the perturbation and $\vec{B}_\mathrm{pert}$ is a unit vector that sets the rotation direction of the perturbation. Its role is to seed a small flavor coherence so that unstable modes can grow during the subsequent evolution. This is particularly useful for MF fast-flavor-instability calculations, where the growth of flavor coherence is one of the central diagnostics.

\paragraph{Trotter time evolution.}
Each one-site time-evolution operator can be written as
\begin{equation}
    U_{1,i}(dt)
    =
    e^{-i h_i dt / \hbar}
    \label{eq:U1}
    \end{equation}
where each $h_i$ is a term in Equation~\ref{eq:qcno_vacuum}. Each two-site interaction can be written as
\begin{equation}
    U_{2,ij}(dt)
    =
    e^{-ih_{ij}dt/\hbar}
    \label{eq:U2}
\end{equation}
where each $h_{ij}$ is a term in Equation~\ref{eq:qcno_self_interaction}. 
For an output interval of $\mathrm{t_{out}}$ divided into $N_\mathrm{steps}$, the code uses $dt=\mathrm{t_{out}}/N_\mathrm{steps}$.

We implement first-order Trotterization as
\begin{equation}
    U_{\mathrm{TEBD1}}(dt)
    =
    \left[\prod_{i=1}^{N}U_{1,i}(dt)\right]\left[\prod_{(i,j)\in \mathcal{P}}U_{2,ij}(dt)\right]
     ,
    \label{eq:unitary_TEBD1}
\end{equation}
where the two products are written in operator order, so the full two-site layer acts first and the one-site layer last. The ordered pair list $\mathcal{P}$ is created with an outer loop over site spacing $d=j-i$. For a given site spacing $d$, site pairs are sorted into even and odd offsets such that within each sub-layer no two operators share a qubit, allowing all operators in a sub-layer to commute and not contribute to Trotter error. This organization allows for all-to-all interactions, and reduces to standard even/odd nearest-neighbor ordering if only $d=1$ operators are active.

We implement second-order Trotterization as
\begin{equation}
    U_{\mathrm{TEBD2}}(dt)
    =
    U^{(\mathrm{R})}_{\mathrm{TEBD1}}\left(\frac{dt}{2}\right)U_{\mathrm{TEBD1}}\left(\frac{dt}{2}\right),
    \label{eq:unitary_TEBD2}
\end{equation}
where $U^{(\mathrm{R})}_{\mathrm{TEBD1}}$ is $U_{\mathrm{TEBD1}}$ with all gates applied in reverse order. The state after a timestep is then 
\begin{equation}
|\psi(t+dt)\rangle = U_{\mathrm{TEBD2}}(dt)|\psi(t)\rangle 
\end{equation}

QCNO evolves particle positions together with flavor. Positions are updated after gates are applied in each Trotter substep according to
\begin{equation}
    \vec x_i\left(t+\frac{dt}{2}\right)
    =
    \vec x_i(t)
    +
    c\,\hat p_i\,\frac{dt}{2} .
\end{equation}
where $dt$ is instead used if the circuit is first order Trotterized (Equation~\ref{eq:unitary_TEBD1}). When using periodic boundary conditions, positions that leave the domain $x\in[0,L)$ are wrapped back into the domain by $x_i \rightarrow x_i \mod L$. The Hamiltonian is rebuilt from the current positions at the start of every substep, so the shape factor in Eq.~\ref{eq:qcno_self_interaction} is evaluated at the substep in which they are applied. Particles are sorted in position order after positions change so spatially local interactions are nearest-neighbor interactions on the device.

\paragraph{Gate implementation.}
Each time-evolution operator is constructed by numerically exponentiating the corresponding Hamiltonian term and appending to the circuit as a dense unitary instruction. That is, we form a circuit out of the unitary operators in Equations~\ref{eq:U1}-\ref{eq:U2} explicitly and allow the Qiskit transpiler to convert them to native gates. Grouping all three axes of each site or pair into one exponential results in a final circuit with smaller trotter error than emitting X, Y, and Z rotations separately, since the Qiskit transpiler attempts to synthesize unitaries into an equivalent sequence of native gates. We find the circuit depth and performance of this approach is superior to emitting explicit rotation gates and either allowing the transpiler to convert to native gates or writing custom Cartan decomposers, since this prevents the unitaries from being optimally synthesized before transpilation, resulting in a much larger circuit depth.

\paragraph{Backend execution.}

The circuit is passed to either a counts-based or direct-state backend dispatcher. In the counts-based route, measurements in the \(X\), \(Y\), and \(Z\) bases are implemented using \(H\), \(S^\dagger H\), and no basis rotation, respectively, followed by computational-basis measurement. Each Pauli component requires an independent set of circuit executions. The qubit-resolved expectation values are obtained from the resulting counts as $
\langle\sigma_{B,i}\rangle=
\left[N_i(+1)-N_i(-1) \right]/N_{\mathrm{shots}},
$
where \(|0\rangle\) and \(|1\rangle\) correspond to eigenvalues \(+1\) and \(-1\), respectively, in the selected basis.

In the direct-state route, the Aer backend saves the full state (either a statevector or density matrix). This route is used for exact state diagnostics, including density matrices, entropy, magic, and Pauli-weight calculations.

The supported backends include two Aer configurations: \texttt{aer\_MF}, which runs as a matrix-product state truncated to bond dimension one, and \texttt{aer\_MB}, an exact statevector simulator. In addition, we use the IBM backends \texttt{FakeManilaV2} (5 qubits, simulated), \texttt{FakeGuadalupeV2} (16 qubits, simulated), and \texttt{ibm\_boston} (156 qubits, hardware). We also use an ideal IonQ cloud simulator (25 qubits, simulated), and IonQ cloud simulator with the \texttt{aria-1} noise model (25 qubits, simulated).

\section{Results}
\label{sec:results}
This work studies whether differences between mean-field (MF) and many-body (MB) inhomogeneous flavor evolution of the spatially inhomogeneous anisotropic two-beam neutrino systems (Sec.~\ref{sec:quantum_framework}) are a feature of correlated dynamics or finite-resource effects.
We frame our discussion around scientific workflow co-design~\cite{barrett2013codesign,shankar2021codesign} focusing on the
application-specific quantities and tradeoffs that connect the physical
problem to near-term and fault-tolerant quantum implementations.
Increasing
the particle number $N$ provides a stronger system-size test but also
increases the number of qubits and active interactions. Reducing the TEBD2
timestep improves numerical resolution but requires more evolution
substeps. The spatial geometry determines which pairs interact and for how
long, while the choice of observable determines the measurement workload.
These application-level choices then propagate into hardware routing,
circuit depth, sampling, and noise, and into the logical circuit size and
rotation-synthesis cost of a fault-tolerant implementation. QCNO keeps the
same physics-defined circuit across these stages so that the physical and
computational tradeoffs can be analyzed consistently.

Hence, we discuss three connected stages of this co-design process.
Application
co-design identifies the physical observables, system sizes, spatial
geometries, and numerical resolution that the circuit must preserve.
Hardware co-design then determines how native gates, connectivity,
compilation, finite sampling, and noise affect those signals and their circuit
cost. Fault-tolerant co-design determines how synthesis accuracy, spatial
geometry, and simulated evolution time set the logical non-Clifford
resources. 

\subsection{Application co-design}
\label{sec:application_codesign}

We simulate the spatially inhomogeneous, anisotropic two-beam neutrino system using both MF and MB evolution under two boundary conditions, giving four simulation configurations. In each case, the $N$ numerical particles are divided equally between oppositely moving beams, with $N/2$ particles initially in electron flavor and $N/2$ in muon flavor. Spatial advection changes the overlap between particle shape functions and, consequently, the active pair interactions.
We consider two geometries: In closed geometry, the two flavor beams initially overlap in the same $1~\mathrm{cm}$ periodic domain. Particles re-enter through the boundaries, allowing repeated interactions between the beams. While in open geometry, the electron-flavor beam initially occupies $0\leq x\leq1~\mathrm{cm}$, while the muon-flavor beam occupies $1\leq x\leq2~\mathrm{cm}$. The beams cross once while their interaction kernels overlap and then separate without re-entry. For each geometry, we perform both MF and MB simulations using classical statevector, allowing us to compare their flavor dynamics under identical spatial conditions and investigate how MB correlations develop during a single crossing versus repeated encounters.

We use application-level tests to investigate two physical signals: polarization dynamics and many-body (MB) structure. First, we increase $N$ in
the closed system and compare MF and MB polarization dynamics to determine
whether the known difference between coherent FFI growth and MB
depolarization persists as the strongest small-$N$ effects decrease. Second,
we use R\'enyi entropy and stabilizer R\'enyi magic to determine what
MB structure accompanies the loss of the one-body signal and how that
structure differs between a single open-boundary crossing and repeated
closed-boundary encounters. The product-formula resolution used for these
calculations is validated independently in
Appendix~\ref{sec:appendix:Numerical_precision}.

The observables have different measurement overheads. The polarization
components used to identify the FFI are sums of single-qubit Pauli
observables and require only global $X$-, $Y$-, and $Z$-basis measurements, independent of particle size $N$.
Entanglement and magic require access to more information
about the MB state. The results below therefore establish both the
one-body and MB signals that the subsequent hardware co-design
analysis must preserve.

\subsubsection{Inhomogeneous fast flavor instability across system size N} \label{sec:inhomo_ffi}
The first application test asks whether QCNO reproduces the known finite-size MF--MB
difference in the Fast Flavor Instability and whether that difference persists as $N$ is increased. Our
previous classical calculations for this same inhomogeneous system reached fully
correlated evolution through $N=24$~\cite{laraib2025two}. Here, the QCNO
quantum-circuit implementation extends the exact MB calculation to $N=30$.

For this test we use the closed geometry defined above and vary $N$ while
keeping the physical model and circuit construction fixed. The MF and MB
calculations use the same Hamiltonian and second-order product formula with
$\mathrm{t_{out}}=5\times10^{-13}~{\rm s}=0.047\mu^{-1}$, $N_{\rm steps}=1$, and
$dt=5\times10^{-13}~{\rm s}=0.047\mu^{-1}$. The MF calculation uses the Aer
matrix-product-state simulator with bond dimension $\chi=1$, which prevents
entanglement, while the MB calculation uses direct Qiskit Aer statevector
evolution and retains the full MB correlations. Because the
observables are evaluated directly from the simulated state, this comparison
contains no finite-shot uncertainty. The electron-beam-averaged
$\langle P_z\rangle$ tracks the loss of the initial flavor polarization,
while $\left|\langle P_x-iP_y\rangle\right|$ tracks the
flavor coherence through the one-site reduced density matrices defined for each site $i$ as
\begin{equation}
\rho_i = \begin{pmatrix}
\rho_{ee,i} & \rho_{e \mu,i} \\
\rho_{\mu e,i} & \rho_{\mu \mu,i}
\end{pmatrix} =
\frac{1}{2}
\begin{pmatrix}
1 + P_{z,i} & P_{x,i} + iP_{y,i} \\
P_{x,i} - iP_{y,i} & 1 - P_{z,i} \\
\end{pmatrix}\,\,,\label{eq:reduced_density_II}
\end{equation}

\begin{figure}
\centering
\includegraphics[width=1.0\linewidth]
{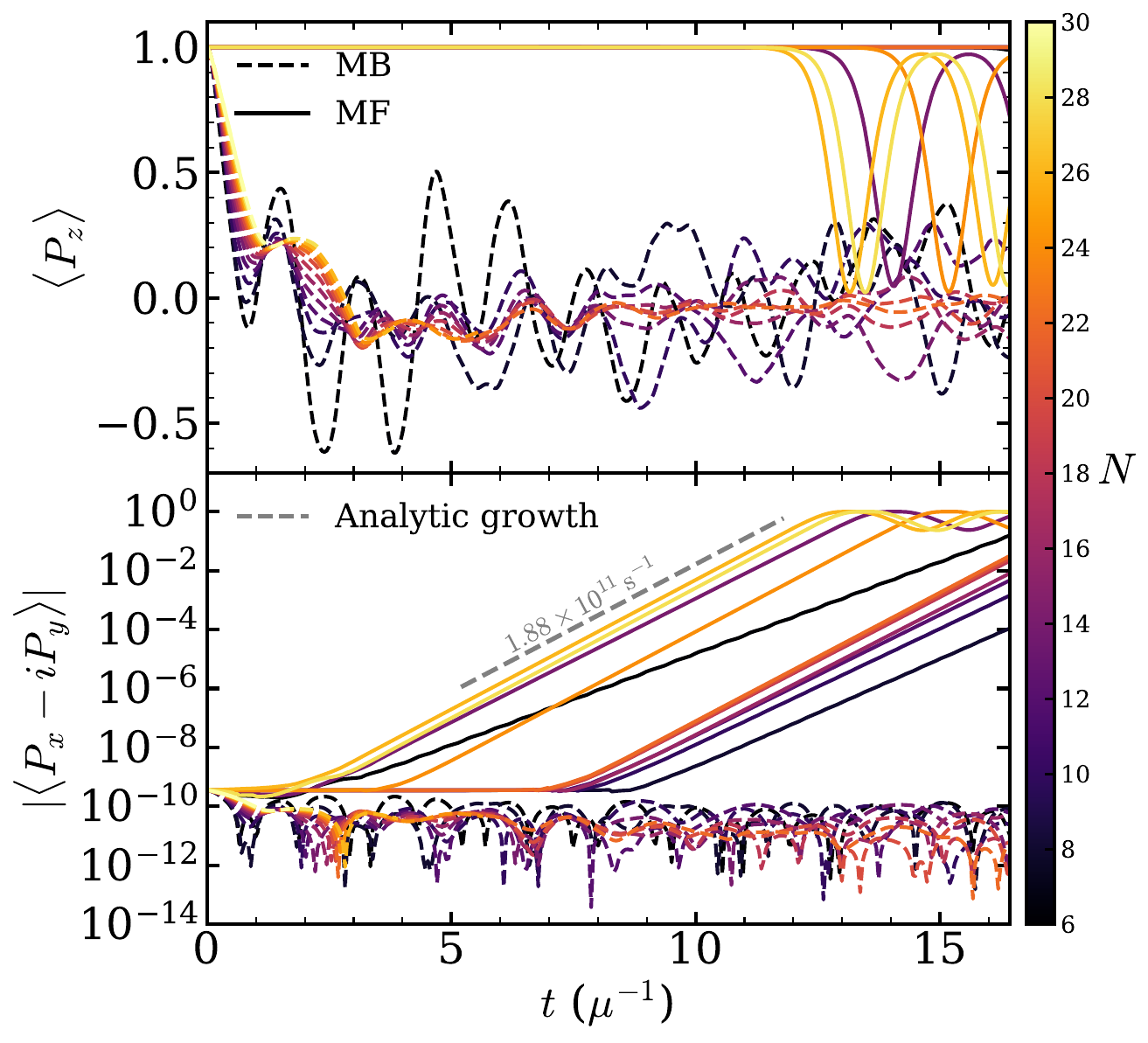}
\caption{
System-size dependence of the spatially inhomogeneous two-beam dynamics for $N=6$--$30$, with $N/2$ particles initially in each flavor beam. The upper panel shows the electron-beam-averaged longitudinal polarization $\langle P_z\rangle$, and the lower panel shows the beam-averaged transverse one-body coherence $\left|\langle P_x-iP_y\rangle\right|$. Colors denote $N$. Solid curves show MF evolution and dashed curves show fully correlated MB evolution from direct Qiskit Aer statevector calculations. Both use the same second-order product formula with $dt=5\times10^{-13}~\mathrm{s}=0.047\mu^{-1}$. The gray dashed line gives the analytic linear-stability growth rate $1.88\times10^{11}~\mathrm{s}^{-1}$, with arbitrary vertical offset. The MF calculation develops exponential transverse growth followed by flavor conversion, while the MB calculation rapidly loses longitudinal polarization without developing the same transverse mode.
}
    \label{fig:scalability_pol_components}
\end{figure}

Figure~\ref{fig:scalability_pol_components} verifies the MF (solid) curves; the fastest unstable mode is picked up through the applied perturbation amplitude of $10^{-2}$ in the transverse |$\langle P_x - i P_y\rangle|$ which exponentially grows showing the Fast Flavor Instability (FFI). Across
these system sizes curves are nearly parallel and approach the analytic
linear-stability growth rate
\(\operatorname{Im}(\Omega)= 2\mu= 1.88\times10^{11}\ {\rm s}^{-1}\)~\cite{chakraborty2016self2,richers2021particle}. Its corresponding $\langle \rho_{ee} \rangle$ transitions from 1 to 0.5 ($\langle P_z\rangle$ from 1 to 0), indicating flavor transformation to a mix of electron and muon neutrinos, before fluctuating back up again. The MF growth rate is similar across system sizes, while the onset and saturation times retain visible finite-$N$ dependence, with $N=6$ showing the largest deviation from the analytical value due to poor spatial resolution.

In contrast, the MB (dashed) curves follow a different path.  Across $N=6$ -- $30$, the $x$ and $y$ components of the polarization vectors remain at the initial scale of $\sim 10^{-10}$ even at late times, while the $z$ components fluctuate near the equilibrated value of 0.  The simultaneous loss of
\(\langle P_z\rangle\) and suppression of
\(\left|\langle P_x-iP_y\rangle\right|\) is therefore consistent with
correlation-driven depolarization and MB dephasing rather than coherent
MF flavor transformation. This MF--MB difference persists through N=30, while the largest finite-$N$ oscillations in the MB $\langle P_z\rangle$ decrease as $N$ increases. Previous calculations for the same inhomogeneous geometry found that MF and MB can approach similar late-time longitudinal polarizations, but on different timescales~\cite{laraib2025two}; the main difference is therefore how, and how quickly the two descriptions lose one-body polarization.

The two calculations allow different information to develop. In MF, each particle remains in a pure one-particle state, so loss of longitudinal polarization is mainly accompanied by coherent rotation into the transverse plane. In MB evolution, entanglement can make the one-particle reduced states mixed, while different transverse phases across particles can also reduce the beam-averaged coherence. One-body polarization can therefore decrease without producing an order-unity transverse signal. Figure~\ref{fig:scalability_pol_components} does not separate these effects, which leads to the next question: if the one-body signal is lost in MB evolution, what MB structure develops instead? We address this next using entanglement entropy and stabilizer R'enyi magic.

\subsubsection{Many-body entanglement and stabilizer R\'enyi magic}
\label{sec:entropy_magic}

The previous section shows that MB evolution loses one-body polarization without
developing the large transverse coherence seen in MF. We now ask what
MB structure accompanies that loss. The second R\'enyi entropy $S_2$
tests whether information initially carried by individual neutrinos becomes
shared through correlations with the rest of the system, while the global
stabilizer R\'enyi magic $M_2$ tests whether the full state also develops
non-stabilizer structure. We evaluate these quantities for $N=6$ and $N=8$
in both the closed and open geometries defined above, allowing us to compare
the structure generated by repeated interactions with that left by a single
beam crossing.

\begin{figure*}
    \centering
    \includegraphics[width=\textwidth]
{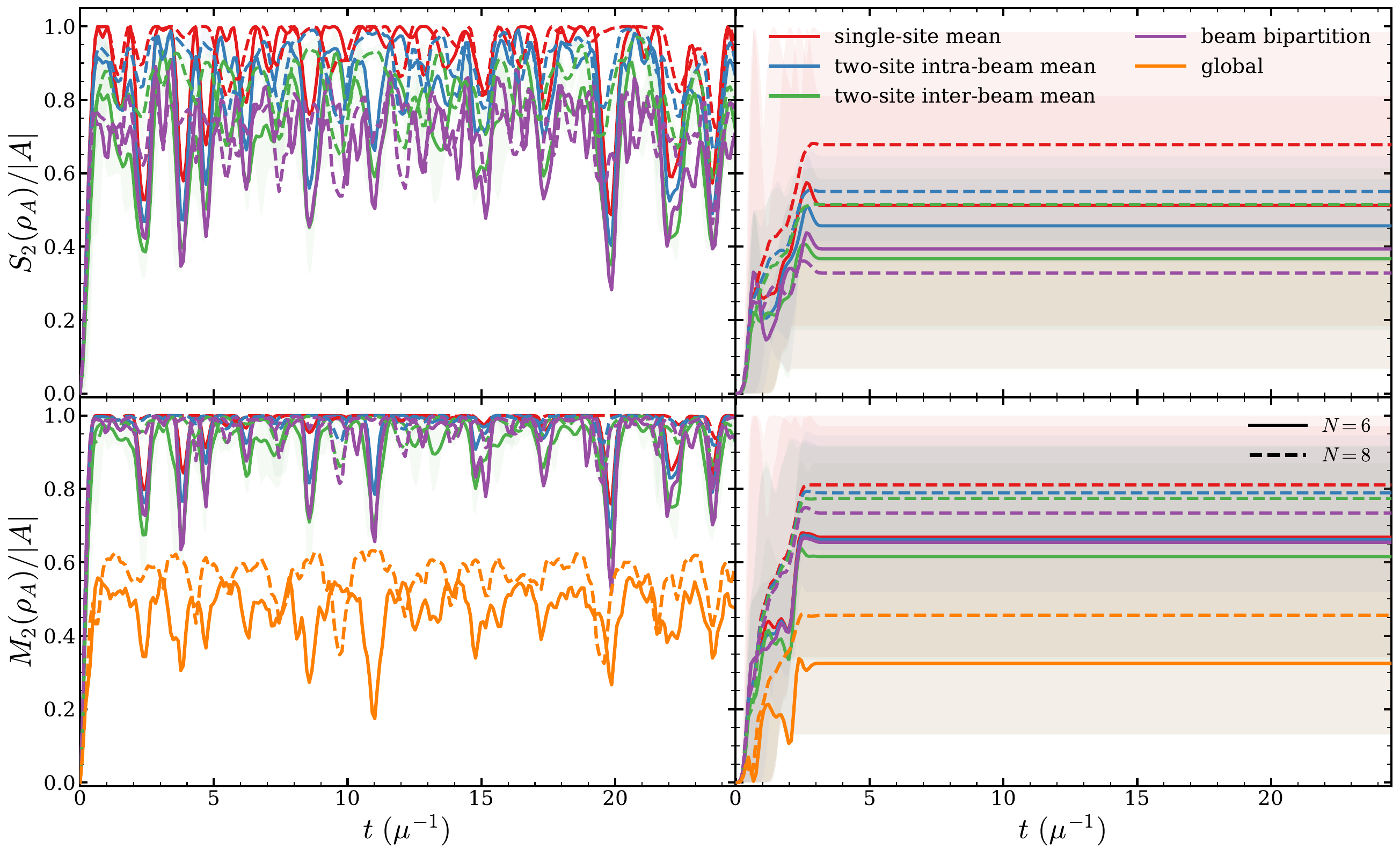}
    \caption{
    Subsystem-size-normalized second R\'enyi entropy and stabilizer R\'enyi
    diagnostics for the inhomogeneous MB evolution with closed periodic
    boundaries (left column) and open boundaries (right column). The upper
    row shows $S_2(\rho_A)/|A|$ and the lower row shows
    $M_2(\rho_A)/|A|$ for $N=6$ (solid) and $N=8$ (dashed). Single-site and two-site curves show averages over the corresponding
    sites or pairs, with shaded regions giving their minimum-to-maximum
    spread. Two-site subsystems are separated into intra-beam and inter-beam
    pairs. The beam bipartition contains all particles belonging to one
    initial flavor beam, while the global $M_2/N$ is evaluated on the full
    pure state. Closed boundaries produce rapid growth followed by continued
    fluctuations as the beams interact repeatedly. Open boundaries produce
    growth during the single beam crossing followed by nonzero plateaus after
    the beams separate. 
    }
    \label{fig:entropy_magic_closed_open_normalized}
\end{figure*}

Figure~\ref{fig:entropy_magic_closed_open_normalized} shows the quantities $S_2(\rho_A)/|A|$ and $M_2(\rho_A)/|A|$, normalized by subsystem $A$ allowing  different subsystems sizes to be compared on a per-qubit basis. Here the subsystem mixedness is characterized by second Rényi entropy 
\begin{equation}
    S_2(\rho_A)
    =
    -\log_2
    \mathrm{Tr}(\rho_A^2).
    \label{eq:renyi2_entropy_results}
\end{equation}
and the non-stabilizer structure (loss of Pauli-basis structure caused by entanglement with $\bar A$) through stabilizer Rényi Magic\cite{froustey2026collective,chernyshev2025quantum}, 
\begin{equation}
    M_2(\rho_A)
    =
    -\log_2
    \left(
        d_A^{-1}
        \sum_{P\in\mathcal P_A}
        |\mathrm{Tr}(\rho_A P)|^4
    \right)\,,
    \label{eq:stabilizer_renyi_magic_results}
\end{equation}
where $d_A=2^{|A|}$ and $\mathcal P$ is the Pauli basis, using reduced density matrices $\rho_A=\mathrm{Tr}_{\bar A}\rho,$ constructed directly from the simulated state
for different subsystem partitions of reduced density matrix. The single-site curves average over $A=\{i\}$ and measure how strongly one neutrino becomes correlated with the remaining system. The two-site curves use $A=\{i,j\}$, where the blue curve averages over pairs in the same beam and the green curve averages over pairs that occupy different beams. Their entropy measures entanglement between the selected pair and the remaining $N-2$ particles, not necessarily entanglement within the pair itself. The beam bipartition contains all $N/2$ particles from one initial flavor beam, and probes entanglement between the complete beams. Global curves are evaluated on $A=\{1,\ldots,N\}$. Since $S_2(\rho_A)$ measures entanglement between $A$ and $\bar A$, for globally pure states, our full ideal state remains pure under unitary evolution, so $S_2(\rho_{\rm global})=0$ at all times and hence is not shown. In contrast, proper subsystems generally have mixed reduced states because they become entangled with their complements. We therefore use their $M_2$ values only as diagnostics of changes in the reduced Pauli-basis structure and do not interpret them in the same way as $M_2$ of the full pure state. The global $M_2$, evaluated on the complete pure state, directly diagnoses the generation of non-stabilizer structure during the many-body evolution. Solid and dashed curves denote $N=6$ and $N=8$, while shaded regions give the minimum-to-maximum spread across the sites or pairs.

The entropy results $S_2(\rho_A)/|A|$ (upper panels) show that the loss of one-body polarization is accompanied
by rapid growth of MB entanglement\cite{froustey2026collective,laraib2025two}. The initial flavor product state
has zero R\'enyi entropy across different subsystem partitions. Under closed boundaries (left panel), the mean single-site
entropy quickly approaches the one-qubit maximum, showing that individual
neutrinos become strongly entangled with the rest of the system. The
two-site intra-beam and inter-beam entropies also become large, and the
beam-bipartition entropy shows that the correlations extend across the two
complete flavor beams. The finite shaded widths show that this structure is
not distributed uniformly: different particles and pairs experience
different interaction histories in the spatially inhomogeneous system. After
the initial rise, the closed-boundary entropies continue to fluctuate because
periodic wrapping produces repeated beam encounters and redistributes the
correlations. In the open geometry (right panel), the same quantities grow while the beams
overlap and then approach nonzero plateaus after the beams separate. A single
crossing is therefore sufficient to leave the final state entangled; repeated
periodic encounters are not required to generate the correlations.

The stabilizer R\'enyi magic diagnostics $M_2(\rho_A)/|A|$ (bottom panels) provide a different view of the same MB
evolution. The normalized $M_2$ values also grow rapidly from their initial
values in both geometries. Under closed boundaries (left panel) they remain time dependent
as the particles repeatedly interact, whereas the open-boundary(right panel) values
approach nonzero plateaus after the interaction ends. The single-site and
two-site reduced density matrices are mixed, so their $M_2$ values
should not be interpreted as direct measures of the non-stabilizer resource of a pure state. Instead, they show
that the reduced states lose the simple Pauli-basis (stablizer) structure of the initial
product state as information is redistributed through the MB system.
The global $M_2/N$, evaluated on the full pure state, gives the direct
state-level indication that the MB evolution also generates non-stabilizer
structure. Entanglement and magic therefore provide complementary
information: the entropy shows that one-body information is shared through
MB correlations, while the global $M_2$ shows that the resulting
MB state also moves away from the stabilizer manifold.

The same qualitative behavior is present for both $N=6$ and $N=8$: rapid
entanglement growth, nonzero global magic, continued evolution under closed
boundaries, and stationary nonzero values after the open-boundary crossing.
The exact normalized values differ between the two sizes, so these results do
not establish convergence or the $N\rightarrow\infty$ scaling of either
quantity. Combined with Sec.~\ref{sec:inhomo_ffi}, however, they give a
consistent picture across the accessible calculations: the suppression of
one-body transverse coherence persists through $N=30$, while the smaller
systems where state-level diagnostics are practical show that this MB
evolution is accompanied by strong entanglement and non-stabilizer structure.

The open and closed geometries therefore provide two different implementation
targets for the remaining QCNO co-design analysis. The open system generates
entanglement and magic during a finite interaction window and then stops
changing once the beams separate, making it a cleaner benchmark for near-term
execution, backend comparison, and noise characterization.
The closed system requires the circuit to preserve changing MB
structure over repeated encounters, making it a more stringent test of circuit depth,
routing, error accumulation, and fault-tolerant resource growth. 
Boundary
conditions are therefore not merely a physical modeling choice; they also
select different circuit-complexity regimes within the same application.

In these ideal
statevector calculations, subsystem mixedness comes only from entanglement
with the remaining simulated particles; on real hardware, environmental
noise can also make reduced states mixed and can therefore mimic or obscure
the physical neutrino signal for polarization, entanglement, and even global magic.

\subsection{Hardware Co-design}
\label{sec:hardware_co_design}

The preceding results assume ideal pure-state evolution, but can we recover these physical signals on realistic quantum hardware? In the noisy intermediate-scale quantum (NISQ) era, how significantly do hardware noise, limited qubit connectivity, and finite sampling affect the measured dynamics? In particular, does the additional routing required
by connectivity-limited superconducting architectures introduce enough
noise to obscure the physical signal, and can all-to-all architectures
reduce this overhead? Can error-mitigation techniques, such as
zero-noise extrapolation (ZNE) and post-processing
denoising~\cite{kemper2024denoising}, recover the underlying dynamics
within practical circuit-depth and sampling limits? In this section,
we examine these questions through hardware co-design, assessing
whether the physics-defined QCNO circuits preserve the target
observables after backend-specific compilation, noise, and finite
sampling are included.

We address this in two steps. First, in
Sec.~\ref{sec:hardware_backend_sensitivity}, we compare the MB polarization
dynamics across backends, system sizes, boundary conditions, routing choices,
and basic error-mitigation strategies.  We then examine architecture-dependent
circuit scaling in Sec.~\ref{sec:hardware_gate_comparison} to determine how the active interaction graph, native gate
set, and hardware connectivity control the one- and two-qubit gate cost of a
product-formula Trotter substep as $N$ increases.

\subsubsection{Noise Sensitivity} 
\label{sec:hardware_backend_sensitivity}
Quantum circuits executed on hardware or in sample-based simulations encounter finite-shot uncertainty and, on noisy backends, errors from imperfect gates and limited qubit coherence. We compare five backends spanning ideal simulation, noisy simulation, and actual hardware execution. Qiskit Aer provides an ideal statevector reference for evaluating finite-shot uncertainty independently of hardware noise. The IonQ ideal simulator tests the same QCNO circuits translated into an all-to-all trapped-ion gate representation, while the IonQ noisy simulator incorporates the Aria gate-noise model. For restricted-connectivity superconducting architectures, we use the IBM Guadalupe noise model and the IBM Boston device for noisy simulation and actual hardware execution, respectively. Unlike the all-to-all trapped-ion architecture, these superconducting backends require additional routing and SWAP-related two-qubit gates to implement nonlocal neutrino pair interactions, increasing circuit depth and potential noise exposure. The IonQ architecture avoids this routing overhead but still accumulates native-gate errors with increasing circuit depth. For all sampled calculations we use
$N_{\rm shots}=1024$ per measurement basis. The statistical uncertainty of $\langle P_z\rangle$ (a single-qubit Pauli expectation) scales as $N_{\rm shots}^{-1/2}$, which we verify
independently using Aer shot scans. We therefore distinguish statistical uncertainty from systematic backend errors: increasing the shot count reduces statistical fluctuations but does not eliminate gate errors, relaxation, readout bias, or routing-induced errors.

We quantify the backend error using the beam-averaged longitudinal
polarization. For backend $b$,
\begin{equation}
\epsilon({\langle P_z \rangle})
=
\sqrt{
\frac{1}{2N_t}
\sum_{\alpha=1}^{2}
\sum_{k=1}^{N_t}
\left[
\langle P_{z,\alpha}^{(b)}(t_k) \rangle
-
\langle P_{z,\alpha}^{({\rm Aer})}(t_k) \rangle
\right]^2
},
\label{eq:backend_pz_rmse}
\end{equation}
where $\alpha=1,2$ labels the two beams, $t_k$ is the $k$th sampled
evolution time, and $N_t$ is the number of samples in the transient
comparison window. This root mean square error formula is appropriate because the measured quantities
are real expectation values rather than full quantum states. It measures the
combined error from incorrect conversion time, damping, oscillatory
structure, and systematic offsets, rather than a
full-state fidelity.

To distinguish connectivity overhead from hardware noise, we test backend-specific layout optimization and three resilience levels ($L0$--$L2$). For the restricted-connectivity IBM Boston and Guadalupe backends, we use custom physical-qubit layouts to reduce routing overhead. We combine these layouts with all three resilience levels for open-boundary runs, while closed-boundary runs use the baseline layout. In contrast, the all-to-all IonQ architecture requires no custom routing. Because its Aria noise model contains no separate readout-noise channel, $L1$ has no effect, although $L2$ still performs zero-noise extrapolation (ZNE)\cite{koenig2024inverted}. Across the backends, $L0$ uses unmitigated sampled expectation values, $L1$ applies per-qubit assignment-matrix readout correction where applicable, and $L2$ combines $L1$ with two-qubit gate folding at noise factors of $1$, $3$, and $5$, followed by quadratic ZNE. Since readout correction and ZNE can produce finite-shot expectation-value estimates outside the physical range $[-1,1]$, we clip the resulting values to this interval.
\begin{figure*}
    \centering
    \includegraphics[width=\textwidth]
    {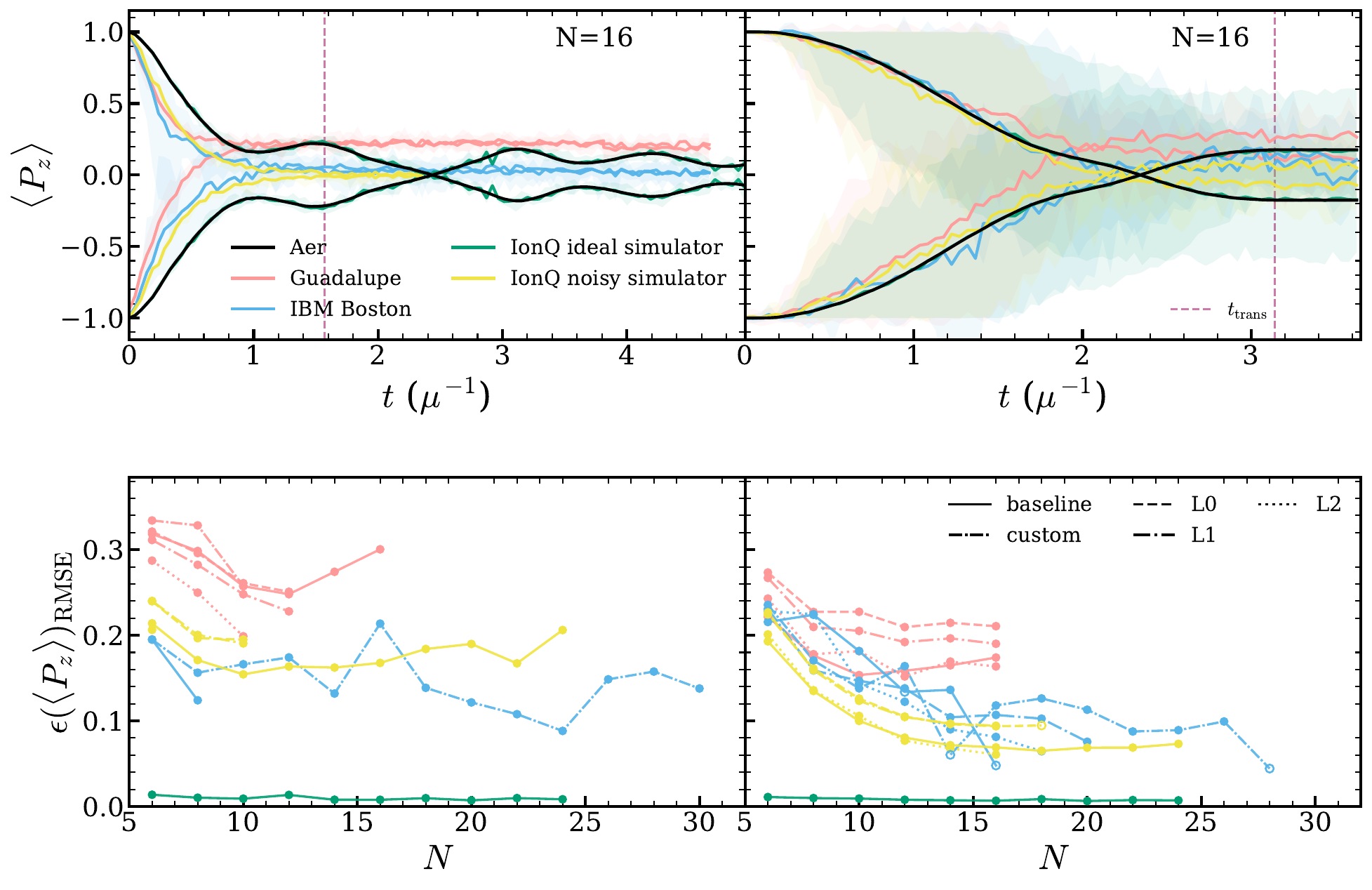}
    \caption{
    Backend sensitivity of the inhomogeneous MB longitudinal polarization for
    closed boundaries (left column) and open boundaries (right column).
    The upper panels show the two beam-averaged $\langle P_z\rangle$
    trajectories at $N=16$, the largest common system size used here for the
    comparison including Guadalupe. Shaded regions give the site-wise
    minimum-to-maximum spread within each beam, while solid lines show beam averages. The vertical (purple) dashed lines mark the time taken for beams to separate in the first pass, i.e., 
    $t_{\rm transient}=  t_{\rm lap}\simeq1.57\,\mu^{-1}$
    for closed boundaries and
    $t_{\rm transient}=2t_{\rm lap}\simeq3.14\,\mu^{-1}$
    for open boundaries. The lower panels show
    $\epsilon({\langle P_z \rangle})_{\mathrm {RMSE}}$ relative to direct Aer as $N$ is increased.
    Colors denote the backend, while line styles distinguish the baseline/default transpilation compared with basic mitigated techniques like 
    custom-layout, and resilience profiles $L0$--$L2$.
    }
    \label{fig:hardware_backend_summary}
\end{figure*}

Figure~\ref{fig:hardware_backend_summary} compares the backend sensitivity of
the inhomogeneous MB polarization for closed boundaries (left
column) and open boundaries (right column). The upper panels show the two
beam-averaged $\langle P_z\rangle$ trajectories at $N=16$, the largest common
system size that  can be run on the Guadalupe architecture. Solid lines show the beam averages, while
the shaded regions give the site-wise minimum-to-maximum spread within each
beam. The vertical purple dashed lines define geometry-based transient windows
($t_{\rm transient}$)
chosen from the spatial dynamics.  In
the closed case, this transient time is chosen as $t_{\rm lap}\simeq1.57\,\mu^{-1}$ which is the time when all sites in the two superimposed beams have fully interacted once before the repeatedly interact in the periodic domain, while in the open case, it is  
$2t_{\rm lap}$, which is the time at which beams separate and no longer interact, allowing fair comparison of accumulated errors in the same single pass of beams across geometries. These accumulated errors are given in the lower panels as 
$\epsilon({\langle P_z \rangle})_{\mathrm {RMSE}}$ relative to direct Aer as $N$ increases. Colors
identify the backend, while line styles distinguish the baseline/default transpilation evolution compared to error-mitigated evolution such as 
custom qubit layout and resilience levels $L0$--$L2$. For each backend and geometry, the upper panel shows the available trajectory with the lowest transient-window RMSE in the corresponding lower panel.
The full $N$-resolved trajectories
are provided in Appendix~\ref{section:backend_open_closed_appendix} (Fig~\ref{fig:hardware_backend_open}). Direct-Aer calculations reach $N=30$, IBM Boston
runs extend to $N=42$, and the IonQ simulator results are shown through $N=24$
because the simulator interface used here is limited to fewer than 25 qubits.

The ideal IonQ simulator closely follows direct Aer in both geometries, with
$\epsilon({\langle P_z \rangle})_{\mathrm {RMSE}}$ remaining near zero throughout the
available $N$ range despite reconstruction from $1024$ shots. The small difference
between these two ideal calculations therefore provides an empirical scale
for the combined finite-shot and ideal native-gate translation error, which
is much smaller than the deviations produced by the noisy backends. This
shows that $1024$ shots are sufficient for the polarization comparison used
here; increasing the shot count would lower the remaining statistical floor
but would not address the dominant backend-dependent errors.

The IonQ noisy simulator damps the polarization magnitude, driving
both beam trajectories toward $\langle P_z\rangle=0$ more rapidly than the
ideal evolution. Its mean RMSE is approximately 18 times higher for closed boundaries and 11 times higher for open boundaries than that of the ideal IonQ simulator. Because the
trapped-ion architecture provides all-to-all connectivity, no routing between
interacting pairs is required, so the additional deviation arises primarily
from the modeled native-gate noise accumulated with circuit depth. The noisy evolution
therefore approaches the same depolarized state as the ideal MB dynamics, but
through gate-induced damping rather than by reproducing the physical
trajectory and timescale. The $L1$ result is unchanged from $L0$, as
expected because the Aria noise model contains no separate readout-noise
channel, while $L2$ tests ZNE. Neither mitigation profile gives a systematic
improvement over the baseline evolution. For open boundaries, $L2$ changes the RMSE by approximately -12\% to +6\% relative to baseline over N=6--16, indicating that mitigation can also increase the error.

Guadalupe gives the largest RMSE among the noisy backends across the available
system sizes in both geometries. It uses a
restricted-connectivity superconducting architecture and therefore incurs
additional routing overhead for the nonlocal interaction graph. Its
trajectories, however, also show a distinct positive
$\langle P_z\rangle$ offset rather than only damping toward zero. Since
$|0\rangle$ contributes $\langle Z\rangle=+1$ and $|1\rangle$ contributes
$\langle Z\rangle=-1$, errors that preferentially relax or assign
$|1\rangle$ as $|0\rangle$ shift the measured polarization in the positive
direction. The observed bias is therefore consistent with asymmetric
relaxation and/or readout error. Custom routing and the $L1$--$L2$ resilience
profiles do not substantially recover the Aer trajectory, and in several
cases give RMSE values comparable to or larger than the baseline result.

Boston uses the same class of restricted-connectivity superconducting
architecture as Guadalupe and therefore carries the same basic routing
penalty for the neutrino interaction graph. Nevertheless, it performs
substantially better than Guadalupe. For the backend properties considered
here, its native-gate error rates are roughly two orders of magnitude
smaller, so circuits of comparable depth accumulate substantially less
hardware error. Its RMSE is consequently comparable to that of the IonQ noisy
simulator over much of their common system-size range, although its
trajectories still show attenuation relative to Aer and the error varies
nonmonotonically with $N$. Neither the custom layout nor the tested resilience
profiles gives a consistent improvement over the default transpiled
evolution, showing that reducing SWAP and routing overhead does not by itself
guarantee better observable accuracy when native-gate noise and decoherence
remain. Boston also supports the largest backend calculations in our sweep,
reaching up to $N=42$ qubits
(Fig~\ref{fig:hardware_backend_open} in Appendix~\ref{section:backend_open_closed_appendix}) which is beyond the direct-Aer statevector range used here. However, this accessible qubit
count does not determine simulation accuracy; the
compiled interaction graph, routing overhead, native-gate quality, and
accumulated two-qubit gate count must all be considered.

Taken together, the backend comparison shows that observable accuracy depends
jointly on architecture, native-gate quality, and mitigation. IonQ avoids the
routing overhead required by the nonlocal interaction graph, whereas Boston
and Guadalupe must map the same interactions onto restricted superconducting
connectivity. Across the noisy backends, Guadalupe gives the largest RMSE (because of its highest native-gate noise),
while Boston and the IonQ noisy simulator are comparable over much of their
common $N$ range. Using the same baseline backend configuration and the common system sizes
available in both geometries, open boundaries reduce the mean RMSE by
approximately $38\%$ for Guadalupe and $48\%$ for the IonQ noisy simulator
relative to closed boundaries; Boston's custom-layout results show
a reduction of about $20\%$ over their common system sizes, although its
baseline results do not show a uniform open--closed trend. The lower errors
in the open geometry are consistent with its finite interaction window, while
the remaining backend differences reflect their respective compilation and
noise characteristics. Likewise, the tested mitigation provides only marginal improvement, and often
none, over the baseline transpiled circuits: custom layouts reduce routing
overhead, $L1$ targets readout bias, and $L2$ applies ZNE, but none removes all
error sources. In particular, ZNE increases
execution cost and estimator variance and is useful only while the physical
signal remains resolvable above the combined sampling and hardware-noise
floor; data-driven denoising faces the same limitation once the signal is
strongly attenuated~\cite{kemper2024denoising}. Because hardware noise also
introduces mixedness unrelated to the physical MB correlations
discussed in Sec.~\ref{sec:entropy_magic}, the polarization dynamics provide
the appropriate first hardware-level benchmark before interpreting more
demanding state-level observables such as entropy or magic. These results motivate
further study of combined and backend-specific error-mitigation strategies for
recovering the physical signal on current NISQ hardware.

Having established the observable-level sensitivity to backend noise,
connectivity, routing, and mitigation, we next ask which circuit features
produce these differences. We therefore turn to the native one- and
two-qubit gate cost and routing overhead of a product-formula substep and
examine how they scale with system size and interaction geometry.

\subsubsection{Architecture-dependent circuit scaling}
\label{sec:hardware_gate_comparison}
As $N$ increases, the hardware cost is set by both the number of active
particle pairs and how those interactions are decomposed and routed on the native hardware
connectivity. Over the $N$ range studied, the Qiskit Aer
runtime scales approximately as $N^2$; for hardware and hardware simulators,
however, wall-clock time also includes queueing, startup, and other
backend-dependent overheads. Native gate counts therefore provide a cleaner
architecture-scaling metric. We examine the native one- and two-qubit gate counts
of a single Trotter substep as a function of $N$, separating the instantaneous
architecture-dependent cost from the additional cost of repeating that
substep over the full evolution. This is particularly relevant for the
inhomogeneous problem, where the active interaction graph changes time-dependently with the
beam overlap.
Circuit depth can be reduced further through circuit compression,
gate scheduling, routing, and backend-native transpilation
~\cite{gibbs2025deep}; here we focus on the architecture-dependent cost already
present in the QCNO implementation.

\begin{figure}
    \centering
    \includegraphics[width=1.0\linewidth]
    {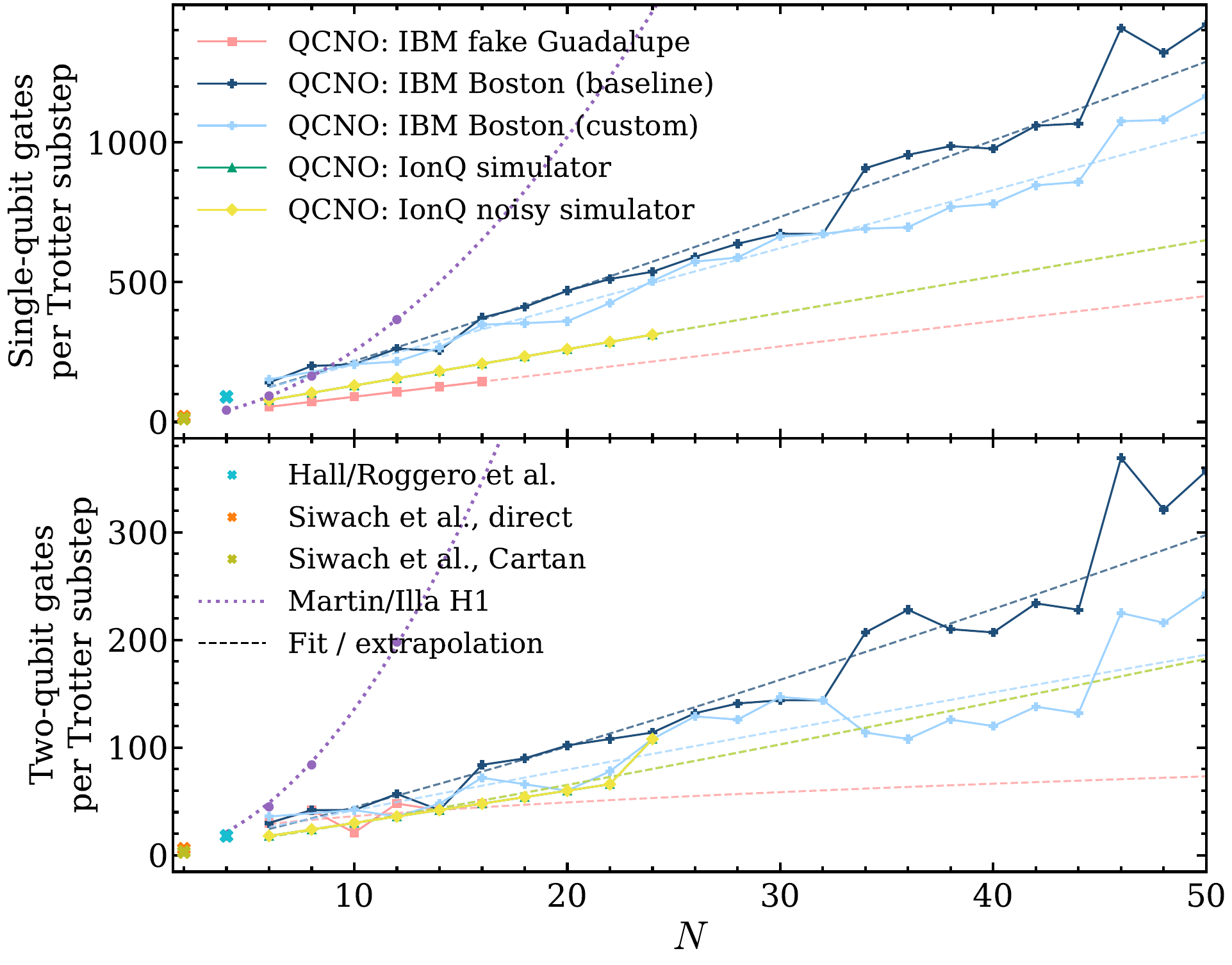}
    \caption{
    Native-gate scaling of one closed-boundary QCNO Trotter substep.
    The upper panel shows the compiled single-qubit gate count and the lower
    panel the corresponding two-qubit entangling-gate count as a function of
    system size $N$. QCNO results are shown for IBM fake Guadalupe, IBM Boston
    with the custom and baseline initial layout, and the IonQ ideal and noisy simulators.
    Hall/Roggero et al.~\cite{hall2021simulation} and Siwach et
    al.~\cite{siwach2023collective} provide fixed small-$N$ reference points,
    while the Martin/Illa H1 results~\cite{illa2023multi} provide a
    homogeneous multi-$N$ comparison. Dashed lines show finite-$N$ fits or
    their extrapolations. The QCNO counts are obtained from the compiled
    $Z$-basis gate-count increment over one product-formula substep.
    }
    \label{fig:one_trotter_gate_comparison}
\end{figure}

Figure~\ref{fig:one_trotter_gate_comparison} shows this one-substep cost for
the closed-boundary calculation. The upper panel gives the backend-native
single-qubit gate count and the lower panel the two-qubit entangling-gate
count. Boston uses the baseline and custom initial layout introduced in the previous
subsection, while the ideal and noisy IonQ simulations have identical gate
counts because the noise model changes the execution rather than the compiled
circuit. The Hall/Roggero et al.~\cite{hall2021simulation} and Siwach et al.~\cite{siwach2023collective} results are included only as fixed
small-$N$ superconducting reference
points (2 and 4 qubits only) rather than scaling datasets. The Martin/Illa H1 results~\cite{illa2023multi} provide the
multi-$N$ trapped-ion comparison. Since the platforms use different native
entangling gates, the lower panel should be interpreted as a comparison of
native two-qubit operation counts rather than as a direct equivalence between
CNOT and $R_{XX}$ gates.

The main result is that the QCNO gate counts grow approximately linearly with
$N$ over the range studied, compared with the approximately quadratic growth
of the homogeneous H1 reference~\cite{illa2023multi}. For H1, a single
evolution step contains $3N+5\binom{N}{2}$ single-qubit gates and
$3\binom{N}{2}$ two-qubit entangling gates, giving the expected
$\mathcal{O}(N^2)$ growth. This follows from the homogeneous isotropic
interaction, where $J_{ij}=1$ and $S(\xi_{ij})=1$, so every particle pair
contributes. In QCNO, however, this shape function kernel $S(\xi_{ij})$ localizes the interaction:
pairs outside the kernel range do not contribute, while active pairs carry
separation-dependent weights. The Trotter substep is therefore a sparse,
nonuniform product over the active pair set rather than an all-pairs product. At fixed density and kernel width, increasing $N$ does not 
increase the number of active neighbors per particle. The total number of active pair rotations therefore scales closer to
$\mathcal{O}(N)$ than $\mathcal{O}(N^2)$ over the accessible range, producing
the near-linear one-substep gate-count trends seen in the compiled QCNO
circuits for the present geometry, cutoff structure, and backend-native gate
decompositions. The fitted curves are empirical finite-$N$ trends for the present geometry,
shape kernel contribution, and backend-native decompositions, not universal complexity
bounds.

The backend curves then show how hardware architecture modifies this near-linear physical scaling. The Hamiltonian interaction graph determines which neutrino
pairs must interact, whereas the hardware coupling graph determines how
costly those interactions are. IonQ provides all-to-all
connectivity, so an active pair can be coupled without additional routing.
Boston and Guadalupe instead map the same interaction graph onto restricted
superconducting connectivity, introducing layout- and routing-dependent
overhead. This produces larger prefactors and
more irregular two-qubit counts even when the underlying active-pair scaling
remains similar.  The custom Boston layout reduces this cost: at
$N=42$, for example, the compiled two-qubit count decreases from $234$ for
the baseline layout to $138$ with the custom layout. The nonmonotonic features
of the superconducting curves are therefore compiler- and routing-dependent
effects rather than changes in the underlying physical scaling.

The gate-count comparison also shows why architecture and gate quality must be
treated separately. Reducing the compiled gate count does not automatically
reduce the observable error: the custom Boston layout lowers routing overhead
but did not produce a systematic reduction in
$\epsilon_{\rm RMSE}(\langle P_z\rangle)$. IonQ avoids a routing penalty but
still accumulates native-gate error with circuit depth, Boston carries a
routing cost but achieves RMSE comparable to noisy IonQ over much of their
common $N$ range, and Guadalupe combines restricted connectivity with larger
backend error. A lower native-gate count therefore reduces one contribution
to the hardware workload, but does not by itself determine fidelity or total
experimental cost, which also depends on gate quality, decoherence, execution
time, and sampling requirements.

Separate open-boundary compilations verified (but not shown here) that the same active-substep
scaling picture holds when the interaction graph changes during the beam
crossing. Before overlap, the interaction-dependent gate cost is nearly
$N$-independent because few interaction terms are active. Near maximum overlap, both
Boston and IonQ recover approximately linear scaling with $N$, consistent
with the active closed-boundary substep. After the beams separate, the
interaction-generated gate increment vanishes. The boundary condition
therefore changes primarily when the active-substep cost is incurred and how
long it persists, rather than the leading $N$ dependence while the
interaction is active. This also explains why the open geometry can show
lower hardware RMSE despite similar active-substep scaling: after a single
crossing the interaction-bearing layers terminate, whereas periodic closed
boundaries produce repeated encounters and continued exposure to hardware
noise. Thus similar per-substep scaling does not imply similar accumulated
circuit cost or hardware fidelity.

Taken together, the hardware-scaling results establish the co-design chain
from spatial structure to near-term execution cost: the geometry determines the active
interaction graph, the backend connectivity and native gate set determine how
that graph is compiled, and the resulting circuit determines the exposure to
hardware noise. The finite-range inhomogeneous interaction gives favorable near-linear
gate growth per active product-formula layer, while the boundary condition
determines how long that cost continues to accumulate. The application-level timestep sets the number of product-formula
repetitions, while backend compilation determines the number and type of
native rotations within each repetition. The relevant hardware tradeoff is
therefore between routing and gate overhead, backend noise, and the statistical noise sampling
required to resolve the target observable as the system size and evolution
time increase. We next ask how these rotations translate into fault-tolerant
logical resources, where arbitrary rotations must be synthesized into
Clifford+$T$ operations and synthesis accuracy introduces an additional
accuracy--cost tradeoff.

\subsection{Fault-tolerant co-design}
\label{sec:fault_tolerant_co_design}
Near-term QCNO execution is limited by accumulated two-qubit error, routing
depth, finite sampling, and other backend-dependent constraints.
Fault-tolerant execution suppresses these physical errors but does not change
the target neutrino dynamics or remove the algorithmic cost of Hamiltonian
simulation. The same MB evolution must still be reproduced at the
required numerical accuracy, while the arbitrary rotations in the
product-formula circuit must now be synthesized into a logical gate set such
as Clifford+$T$. This introduces synthesis error and a corresponding
non-Clifford resource cost, so product-formula and synthesis errors must be
controlled within a common accuracy budget. We therefore combine the
active-pair structure established above with rotation-synthesis accuracy to
determine the logical non-Clifford resources required by QCNO. We do not
assign a universal system size at which fault-tolerant execution becomes
preferable to NISQ execution, since that crossover depends on hardware error
rates, connectivity, achievable depth, measurement cost, and the chosen
error-correction architecture.

We retain the product-formula framework used throughout QCNO because it
provides a qubit-efficient, direct circuit representation of the Hamiltonian
that is applicable to both near-term and fault-tolerant implementations
~\cite{clinton2021hamiltonian,bosse2025efficient}. This also keeps the
physical, near-term, and fault-tolerant resource analyses directly
comparable. The timestep-convergence study in
App.~\ref{sec:appendix:Numerical_precision} shows that the polarization
observables are converged at the production timestep
$dt=5\times10^{-13}\,\mathrm{s}$. 
Other fault-tolerant Hamiltonian-simulation approaches,
including block encoding, linear combinations of unitaries, and quantum
signal processing, offer different resource tradeoffs
~\cite{low2017optimal,bauer2023hep}; here we restrict the analysis to the
QCNO product-formula circuits.

We address three connected questions: how the logical non-Clifford cost
depends on the rotation-synthesis tolerance, how the spatial geometry and
boundary conditions determine the accumulation of that cost with simulated
time, and what these results imply for the transition from NISQ to
fault-tolerant execution. The stabilizer R\'enyi magic in
Sec.~\ref{sec:entropy_magic} establishes that the dynamics generate
non-Clifford state structure, but it does not determine the $T$-gate cost of
a particular circuit. That cost depends on the number of retained circuit
rotations and the accuracy with which they are synthesized. We therefore
quantify logical $T$-gate resources in this section; code distance,
magic-state factories, physical-qubit overhead, and full logical runtime
depend on the chosen error-correction architecture and are beyond the present
estimates.\cite{hirano2024magicpool,jacinto2026exploring, wills2025constant,acharya2024quantum}.

\subsubsection{Synthesis accuracy and non-Clifford cost}
\label{sec:ft_synthesis_accuracy}

At the fault-tolerant level, we characterize the circuit in terms of logical
gate counts rather than the device-native physical gates used in the NISQ
analysis. We use the standard Clifford+$T$ gate set. The Clifford operations,
generated by $H$, $S$, and CNOT together with Pauli gates, map Pauli
operators to Pauli operators under conjugation. Adding the non-Clifford
$T$ gate and its inverse $T^\dagger$ makes this discrete gate set universal.
In fault-tolerant architectures, logical $T$ gates are typically implemented
by consuming high-fidelity magic states, whose preparation and distillation
can dominate the non-Clifford resource overhead. A complete architecture
estimate would additionally require the error-correcting code and code
distance, magic-state factories, physical-qubit overhead, and logical runtime\cite{o2017quantum, hirano2024magicpool,jacinto2026exploring, wills2025constant,acharya2024quantum}. 

QCNO evaluates the non-Clifford workload using two complementary compilation
diagnostics, both transpiled at optimization level $1$, which applies
lightweight circuit simplification and basic gate cancellation:
\begin{equation}
\begin{aligned}
\mathcal B_{\rm disc}
&=
\{
\mathrm{CNOT},H,S,S^\dagger,X,\mathrm{measure},T,T^\dagger
\}, \\
\mathcal B_{R_Z}
&=
\{\mathrm{CNOT},H,S,S^\dagger,X,\mathrm{measure},R_Z\}.
\label{eq:ft_resource_bases}
\end{aligned}
\end{equation}
The first diagnostic is the sum of $T$ gates read from the circuit compiled directly into the $\mathcal B_{\rm disc}$ (i.e., discrete Clifford+$T$) basis:
\begin{equation}
N_T^{\rm raw}
=
N_T+N_{T^\dagger}.
\label{eq:raw_t_count_methods}
\end{equation}
Because this count depends on the compiler, target basis, and optimization
workflow, it should be interpreted as a compilation-level resource diagnostic
rather than a unique or optimal logical $T$ count~\cite{Mooney_2021}.

We also compare this to an analytic estimate of the $T$ gates required to synthesize the $R_Z$ rotations in a circuit compiled with the $\mathcal{B}_{R_Z}$ basis. In a fault-tolerant implementation, these
continuous rotations must be synthesized, meaning approximated by finite
sequences of discrete Clifford+$T$ operations; higher synthesis accuracy
therefore requires longer gate sequences and a larger $T$ count. The TEBD2 evolution contains Pauli exponentials
of the form $e^{-i\theta P}$, where $P$ is a Pauli string. Up to Clifford
basis changes and parity-computation operations, each such exponential can be
reduced to a single arbitrary $R_Z$ rotation,
\begin{equation}
e^{-i\theta P}
=
C_P^\dagger R_Z(2\theta)C_P ,
\label{eq:pauli_rotation_to_rz_methods}
\end{equation}
where $C_P$ is composed of Clifford operations such as $H$, $S$, and CNOT
that map $P$ to a single-qubit $Z$ operator, and
$R_Z(\phi)=e^{-i\phi Z/2}$. These Clifford operations contribute to the
logical gate count and circuit depth but not directly to the non-Clifford
$T$ count. For an $R_Z$ rotation approximated to tolerance
$\varepsilon_{\rm syn}$, the required number of $T$ gates is estimated as~\cite{RossSelinger2016}
\begin{equation}
N_T^{(R_Z)}(\varepsilon_{\rm syn})
\simeq
-3\log_2\left(\varepsilon_{\rm syn}\right),
\label{eq:single_rz_t_cost_methods}
\end{equation}
For a circuit containing $N_{R_Z}$ explicit
$R_Z$ rotations, we estimate the total synthesis contribution as
\begin{equation}
N_T^{\rm est}
=
N_{R_Z}N_T^{(R_Z)}(\varepsilon_{\rm syn}),
\label{eq:rz_estimated_t_count_methods}
\end{equation}
where $N_{R_Z}$ is the number of arbitrary $R_Z$ rotations requiring
synthesis.

\begin{figure}
    \centering
    \includegraphics[width=1.0\linewidth]{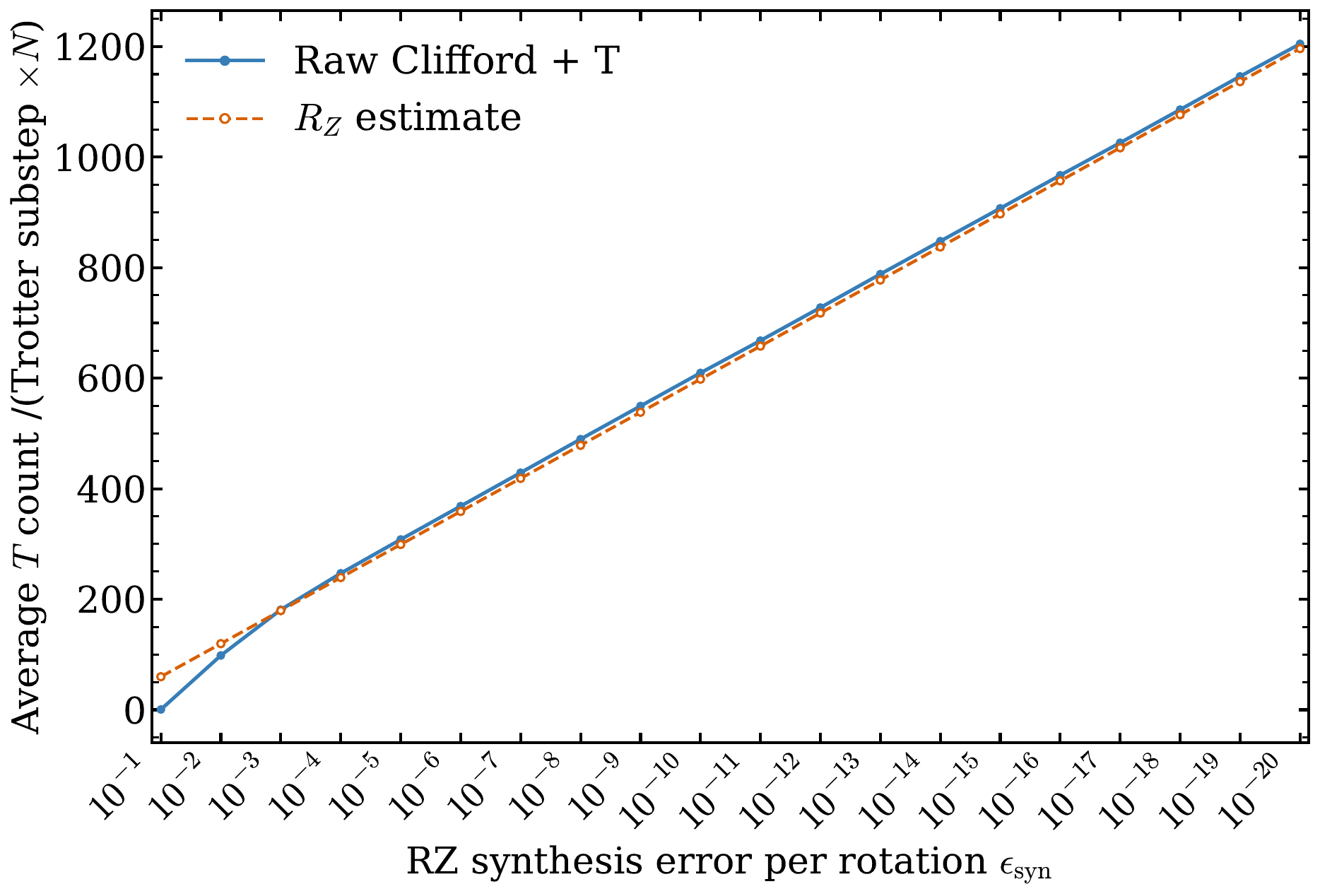}
    \caption{
    Average fault-tolerant $T$ count per Trotter substep, normalized by
    system size $N$, as a function of the per-rotation synthesis tolerance
    $\epsilon_{\rm syn}$ for the closed-boundary inhomogeneous MB circuits compiled . The solid curve shows the directly compiled
    Clifford+$T$ count, $N_T+N_{T^\dagger}$, obtained from Qiskit's
    Clifford+$T$ synthesis pass, while the dashed curve shows the analytical
    $R_Z$-synthesis estimate of
    Eq.~\eqref{eq:rz_estimated_t_count_methods}. Both increase approximately
    as $\log(1/\epsilon_{\rm syn})$ as the requested synthesis accuracy is
    tightened. Normalization by $N$ exposes the approximately linear
    system-size scaling of the per-substep non-Clifford workload over the (50 qubits)
    range studied. All circuits are transpiled at optimization level $1$.
    }
    \label{fig:rz_t_slope_epsilon}
\end{figure}
Figure~\ref{fig:rz_t_slope_epsilon} validates the analytical $R_Z$-synthesis
estimate against the $T+T^\dagger$ count obtained directly from the
Clifford+$T$ compiler. Both increase approximately linearly with
$-\log\epsilon_{\rm syn}$, as expected from
Eq.~\eqref{eq:single_rz_t_cost_methods}, showing that tighter synthesis
accuracy increases the non-Clifford gate requirement. After normalizing the
average $T$ count per Trotter substep by $N$, the $R_Z$-based estimate varies
by only about $0.45\%$ across the system sizes studied, while the directly
compiled Clifford+$T$ counts remain within about $5.1\%$ of the same
linear-in-$N$ scaling, except at $\epsilon_{\rm syn}=10^{-1}$ where the small
discrete gate counts produce larger relative variation. At the production
choice $\epsilon_{\rm syn}=10^{-7}$, the directly compiled and estimated
costs are approximately $429$ and $418$ $T$ gates per Trotter substep per
qubit, respectively, differing by about $2.4\%$. 

To quantify how the per-rotation synthesis error accumulates across the
circuit, we write the exact and synthesized TEBD2 implementations as
\begin{equation}
U_{\rm TEBD2}
=
U_{N_{R_Z}}\cdots U_1,
\qquad
\widetilde U_{\rm TEBD2}
=
\widetilde U_{N_{R_Z}}\cdots\widetilde U_1 ,
\label{eq:synthesized_product_ft_methods}
\end{equation}
where $N_{R_Z}$ is the number of arbitrary $R_Z$ rotations requiring
synthesis, $U_j$ is the exact circuit factor associated with the $j$th
rotation, and $\widetilde U_j$ is its Clifford+$T$ approximation. The
Clifford operations contained in $U_j$ and $\widetilde U_j$ are identical;
the difference between the two factors arises only from the finite-precision
synthesis of the non-Clifford rotation. We quantify this difference using the
operator $2$-norm,
$\|A\|_2=\sup_{\|\psi\|_2=1}\|A\psi\|_2$, and assign the same synthesis
tolerance to each rotation,
\begin{equation}
\left\|
U_j-\widetilde U_j
\right\|_2
\leq
\epsilon_{\rm syn},
\label{eq:single_rotation_synthesis_error_ft_methods}
\end{equation}
Because the remaining factors are unitary, the triangle inequality gives the
conservative accumulated bound
\begin{equation}
\left\|
U_{\rm TEBD2}
-
\widetilde U_{\rm TEBD2}
\right\|_2
\leq
\sum_{j=1}^{N_{R_Z}}
\left\|
U_j-\widetilde U_j
\right\|_2
\leq
N_{R_Z}\epsilon_{\rm syn}.
\label{eq:synthesis_error_bound_ft_methods}
\end{equation}
This bound assumes the individual synthesis errors could add maximally;
in practice, errors from different rotations need not accumulate coherently.

However, synthesis error is only one error contributing to the accuracy of a fault-tolerant circuit. The second error source is the TEBD2 splitting approximation. These product-formula
errors arise from noncommutativity among Hamiltonian terms, with symmetric
second-order formulas governed by the corresponding higher-order
nested-commutator structure~\cite{Childs2021Trotter}. Because the interaction
Hamiltonian changes with advection, we evaluate this error directly rather
than through an analytic commutator bound. For each piecewise-frozen
Hamiltonian $H_k$, we compare the exact substep propagator
$U_{{\rm exact},k}=\exp(-iH_kdt/\hbar)$ with the corresponding
symmetric TEBD2 propagator $U_{{\rm TEBD2},k}$ and define the local
splitting error
\begin{equation}
e_k
=
\left\|
U_{{\rm exact},k}
-
U_{{\rm TEBD2},k}
\right\|_2 .
\label{eq:tebd2_local_error_results}
\end{equation}
Each $e_k$ therefore measures only the error generated during one Trotter
substep and does not include errors accumulated at earlier times.

\begin{figure}
    \centering
    \includegraphics[width=1.0\linewidth]
    {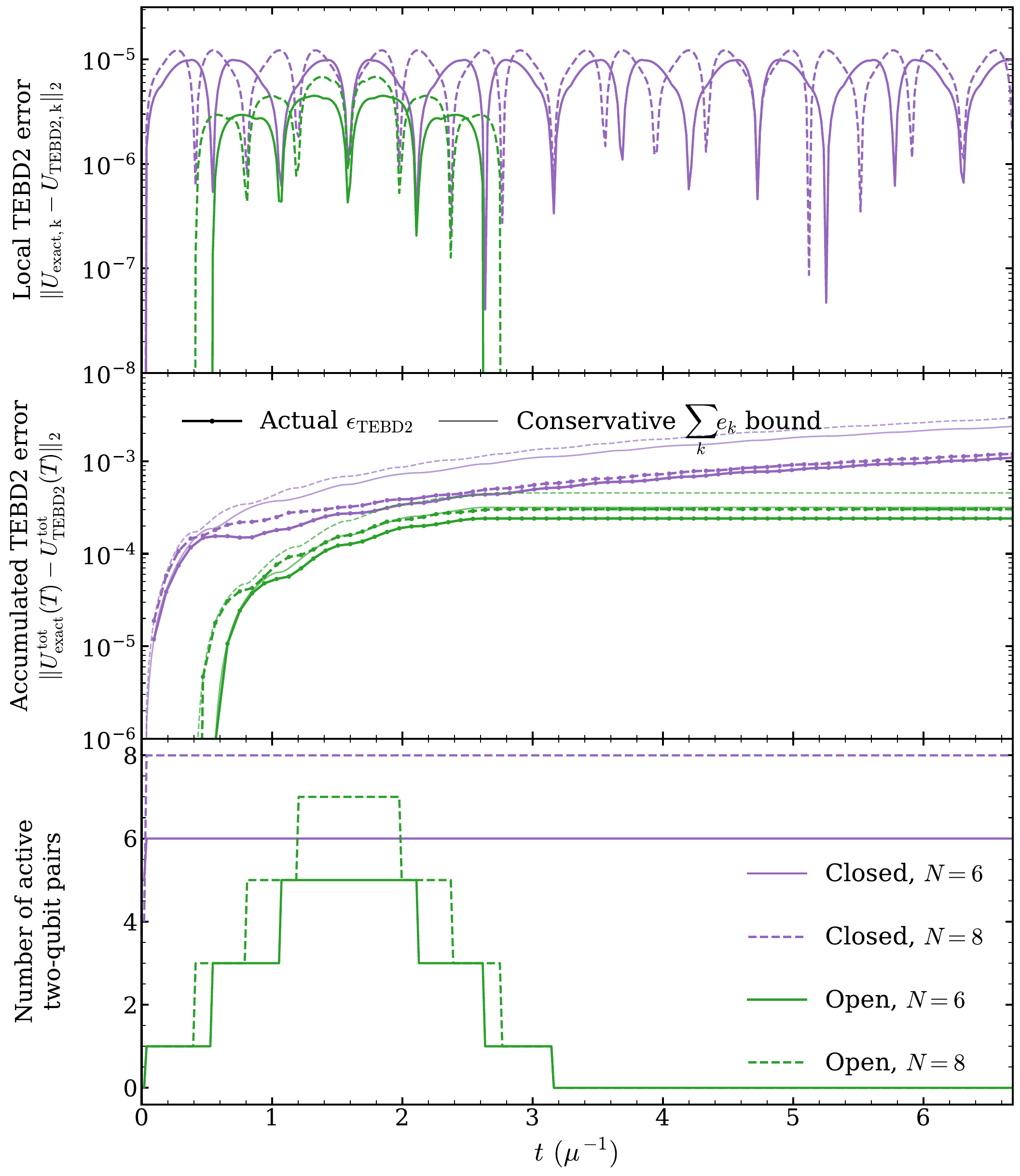}
    \caption{
    TEBD2 product-formula error and active interaction pairs for the closed and open
    $N=6$ and $N=8$ inhomogeneous MB circuits with timestep $dt=2\times10^{-13}\ {\rm s}=0.019 \mu^{-1}$. The upper panel shows the local
    operator-norm error
    $e_k=\|U_{{\rm exact},k}-U_{{\rm TEBD2},k}\|_2$
    for each Trotter substep. The middle panel shows the accumulated TEBD2
    error
    $\epsilon_{\rm TEBD2}(T)
    =\|U_{\rm exact}^{\rm tot}(T)-U_{\rm TEBD2}^{\rm tot}(T)\|_2$
    together with the conservative bound $\sum_k e_k$. The lower panel
    shows the number of active two-qubit interaction pairs at each Trotter
    substep. Purple and green curves denote the closed and open geometries,
    respectively, while solid and dashed curves denote $N=6$ and $N=8$.
    }
    \label{fig:tebd2_operator_error}
\end{figure}

In
Fig.~\ref{fig:tebd2_operator_error}, purple and green curves denote the
closed and open geometries, respectively, while solid and dashed curves
denote $N=6$ and $N=8$ small system sizes we considered for this TEBD2 cost analysis. The upper panel shows $e_k$, the middle panel the
accumulated TEBD2 error and its conservative bound, and the lower panel the
number of active two-qubit pairs showing how the spatial interaction
structure controls where TEBD2 error is generated. The closed geometry
keeps all qubits active, whereas the open
geometry activates interactions only while the initially separated beams
approach and cross before the active-pair count returns to zero. During
active interactions, $e_k$ is typically of order $10^{-6}$--$10^{-5}$.
Its detailed variation does not follow the pair count alone because it also
depends on the instantaneous interaction strengths and noncommutativity of
the active Hamiltonian terms. To quantify the net error through Trotter
substep $K$, we construct
$U_{\rm exact}^{\rm tot}(T)=U_{{\rm exact},K}\cdots U_{{\rm exact},1}$
and
$U_{\rm TEBD2}^{\rm tot}(T)=U_{{\rm TEBD2},K}\cdots
U_{{\rm TEBD2},1}$ and define
\begin{equation}
\epsilon_{\rm TEBD2}(T)
=
\left\|
U_{\rm exact}^{\rm tot}(T)
-
U_{\rm TEBD2}^{\rm tot}(T)
\right\|_2
\leq
\sum_{k=1}^{K}e_k .
\label{eq:tebd2_actual_error_results}
\end{equation}
Here $\epsilon_{\rm TEBD2}(T)$ is the directly calculated accumulated
splitting error, while $\sum_k e_k$ is its conservative triangle-inequality
bound; local errors from different substeps need not add coherently. In the
closed geometry, the accumulated error continues to evolve as interactions
recur. In the open geometry, it rises during the beam crossing and
approaches a plateau once the interaction window closes. 

This calculation isolates the TEBD2 splitting error obtained when the
time-dependent Hamiltonian is treated as constant within each Trotter
substep, with a fixed Hamiltonian $H_k$ used over substep $k$.  It therefore
measures the precision of the TEBD2 factorization for the chosen dt,
but does not include the separate time-discretization error introduced by
freezing the continuously varying position-dependent interaction within each substep.
For each piecewise-fixed Hamiltonian $H_k$, the local error of the
symmetric second-order product formula scales as
\begin{equation}
e_k \le c_k\,(\mu\,dt)^3
\end{equation}
for second-order Trotterization, where $c_k$ is determined by nested commutators among the
noncommuting Hamiltonian terms active during substep $k$. While the Trotterization is second order, the coefficients of the terms in the Hamiltonian are held constant during each substep as in a forward-Euler integration scheme. If advection errors dominate, we expect the integration to be first order, but where advection is not important (i.e., Hamiltonian coefficients do not change quickly) we expect second order convergence. The total operator error then scales with the convergence order $p$ (between 1 and 2 for our model depending on the circumstance).
Using Eq.~\eqref{eq:tebd2_actual_error_results}, the accumulated TEBD2 splitting error satisfies
\begin{equation}\begin{aligned}
\epsilon_{\rm TEBD2}(T)
&\le \sum_{k=1}^{K} e_k \le \sum_{k=1}^{K} c_k\,(\mu\,dt)^{1+p}\\
&\le c_{\max} \mu^{1+p} T\,dt^p,
\end{aligned}
\end{equation}
where $K=T/dt$ and $c_{\max}=\max_k c_k$. Therefore, when the
TEBD2 splitting error is the dominant numerical contribution,
achieving a target error $\epsilon$ requires
\begin{equation}
\label{eq:ntrotter_estimate}
K \gtrsim
\mu\,T\left[\frac{c_\mathrm{max}\mu\,T}{\epsilon}\right]^{1/p}
\end{equation}
so the number of Trotter substeps scales as
$K\propto\epsilon^{-1/p}$, as expected for a $p$-order product
formula. This scaling refers specifically to the TEBD2 splitting
error; the separate error associated with freezing the
time-dependent Hamiltonian coefficients within each substep can
reduce the overall convergence order. Estimating the single-step ($K=1$, $T=dt=0.019\mu^{-1}$) error as $\epsilon=10^{-5}$ from Fig.~\ref{fig:tebd2_operator_error}, we estimate $c_\mathrm{max}\approx0.028$ for p=1 or $c_\mathrm{max}\approx1.5$ for p=2. Note that these calculations use a smaller $dt$ than production calculations to ensure that $\epsilon_{\mathrm {TEBD2}}$ serves as a stringent limit.

With the synthesis and TEBD2 errors now defined separately, their joint
accuracy budget is
\begin{equation}
\epsilon_{\rm tot}(T)
=
\epsilon_{\rm TEBD2}(T)
+
\epsilon_\mathrm{syn}(T)
\label{eq:tebd2_synthesis_total_error_results}
\end{equation}
where $\epsilon_\mathrm{syn}(T)=\varepsilon_{\rm syn} N_{R_Z}(T)/N$.

A useful synthesis scale is obtained by requiring the conservative
synthesis contribution to be comparable to the measured TEBD2 error,
\begin{equation}
\epsilon_\mathrm{syn}(T)\simeq \epsilon_\mathrm{TEBD2}(T),
\label{eq:matched_synthesis_tolerance_results}
\end{equation}
At the common comparison time, the error-matching condition gives
$\epsilon_{\rm syn}^{\rm match}\simeq8.5\times10^{-8}$ for $N=6$ and
$7.0\times10^{-8}$ for $N=8$, identifying a useful synthesis tolerance scale of order
$10^{-7}$. The larger-$N$
resource calculations use the same Hamiltonian construction, timestep, and
synthesis model, and show the same approximately linear rotation-workload
scaling with $N$, so we use $\epsilon_{\rm syn}=10^{-7}$ as a representative
tolerance for the larger-system resource estimates as well. This choice
should be interpreted as an application-level extrapolation rather than an
independently verified error-balanced tolerance at every $N$.

The number of Z rotations in a single TEBD2 timestep using closed boundaries is $N_{R_Z}(dt)=6N$ in the circuits compiled for this problem. The number of Z-rotations for a given total time $T$ is then $N_{R_Z}(T) = 6NK$ and the allowed synthesis tolerance is $\varepsilon_\mathrm{syn}=\epsilon/6K$. The required synthesis tolerance is then determined by plugging these scalings into Eq.~\ref{eq:matched_synthesis_tolerance_results}:
\begin{equation}
\label{eq:esyn_final_estimate}
\varepsilon_\mathrm{syn} \simeq \frac{1}{6c_\mathrm{max}^{1/p}} \left(\frac{\epsilon}{\mu T}\right)^{1+\frac{1}{p}}\,.
\end{equation}
Equations~\ref{eq:rz_estimated_t_count_methods}, \ref{eq:ntrotter_estimate}, and \ref{eq:esyn_final_estimate} estimate the resources required for a simulation of duration $T$ and error $\epsilon$ for closed-boundary calculations. Since open-boundary calculations are cheaper, these serve as upper bound estimates for either geometry.

\subsubsection{Geometry-dependent accumulation of \(T\)-gate resources}
\label{sec:ft_boundary_geometry}

\begin{figure}
    \centering
    \includegraphics[width=1.0\linewidth]
    {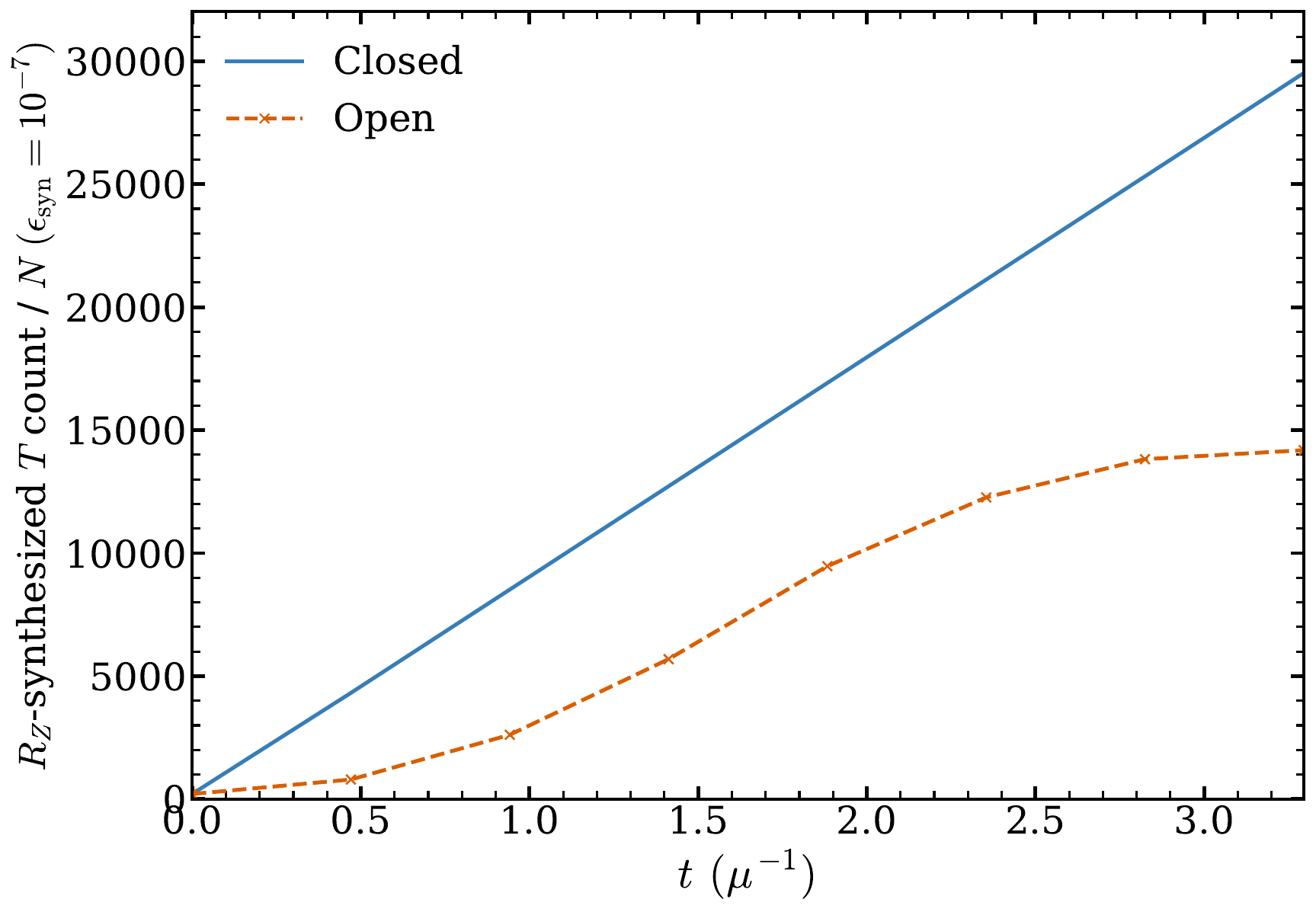}
    \caption{
    Accumulated $R_Z$-synthesized $T$ count per N particles, for the
    closed (solid blue) and open(dashed orange) inhomogeneous MB circuits through $N=50$, evaluated with
    $dt=5\times10^{-13}\ {\rm s}=0.047\mu^{-1}$ and
    $\epsilon_{\rm syn}=10^{-7}$. Normalization by $N$ causes the
    system-size curves to nearly collapse, showing approximately linear
    scaling of the accumulated $T$ count with $N$ over the range studied.
    Closed boundaries produce nearly linear resource accumulation in time,
    whereas the open-boundary cost grows nonlinearly during the finite beam
    crossing and approaches a plateau as the beams separate.
    }
    \label{fig:rz_t_count_boundary_comparison}
\end{figure}

Geometry also controls how the non-Clifford workload accumulates in time.
The analysis in Sec.~\ref{sec:ft_synthesis_accuracy} used the closed periodic
geometry, where repeated interactions produce an approximately stationary
rotation workload. In the open geometry, the number of active interactions
changes throughout the single beam crossing, so we compare the accumulated
$R_Z$-synthesized $T$ count directly. 
Figure~\ref{fig:rz_t_count_boundary_comparison}
shows this quantity normalized by $N$ for system sizes through $N=50$, using
the production timestep $dt=5\times10^{-13}\ {\rm s}=0.047\mu^{-1}$ and the
representative synthesis tolerance $\epsilon_{\rm syn}=10^{-7}$ motivated by
the error-budget analysis above.

The normalized curves show that the accumulated synthesis workload is
approximately linear in system size over the range studied. For closed
boundaries, the normalized curves differ by at most about $3.0\%$, while the
open-boundary curves differ by at most about $6.7\%$, with the largest
deviations occurring at early times when only a small and changing set of
particle pairs is interacting. Thus, system size primarily sets the overall
resource scale, while the boundary geometry determines how that resource
accumulates in time. Periodic wrapping in the closed geometry repeatedly
restores interacting pairs and produces nearly linear growth of the
accumulated $T$ count, whereas the open geometry grows nonlinearly; slowly as the beams
begin to overlap, more rapidly during the crossing, and approaches a plateau
after separation when no new interaction-generated rotations are produced.
This finite interaction window is also consistent with the plateau behavior
of the open-boundary entanglement and magic diagnostics in
Sec.~\ref{sec:entropy_magic}, although these state-complexity measures and
the circuit $T$ count are not expected to be quantitatively proportional. 
At $\epsilon_{\rm syn}=10^{-7}$, the open-boundary circuits reach a favorable total synthesis cost of $1.4\times10^4$ \(T\) gates, about 2 times less than closed boundaries per system size, over the interval shown.

\subsubsection{Implications for the NISQ-to-fault-tolerant transition}
\label{sec:ft_nisq_transition}

The hardware and fault-tolerant co-design results together show that favorable
circuit scaling does not by itself guarantee accurate quantum execution.
Although the backend-transpiled QCNO circuits exhibit approximately linear
finite-size gate scaling with $N$ over the range studied, the noisy
implementations produce polarization errors of
$\epsilon(\langle P_z\rangle)_{\rm RMSE} \sim 0.1$--$0.3$. These deviations
arise from accumulated backend-dependent gate noise, decoherence, routing,
and readout effects, and cannot be removed simply by increasing the number of
shots. This hardware-error scale is much larger than the numerical TEBD2
splitting error resolved in the fault-tolerant analysis,
$\epsilon_{\rm TEBD2}\sim10^{-3}$. The present results therefore motivate
error-corrected execution if the numerical accuracy of the underlying
MB calculation is to be preserved beyond the accuracy reached by the
NISQ implementations studied here\cite{acharya2024quantum}.

The fault-tolerant resource target follows directly from the application
error budget. Balancing the accumulated synthesis error against the measured
TEBD2 error gives the representative tolerance
$\epsilon_{\rm syn}=10^{-7}$, corresponding to approximately $70$ $T$
gates per synthesized $R_Z$ rotation and fixing the total non-Clifford
workload of the circuit. A fault-tolerant implementation must then suppress
the remaining hardware-induced logical failures below a separate reliability
target. If a total logical-failure probability
$\epsilon_{\rm L}\sim10^{-3}$ is allowed and independent logical $T$-gate
faults dominate this budget, the required per-$T$ logical error rate is
approximately $p_T\sim\epsilon_{\rm L}/N_T$. Thus, for a circuit with
$N_T\sim10^4$ logical $T$ gates, one obtains
$p_T\sim10^{-7}$ per qubit. This is an application-driven logical-reliability target,
not a complete fault-tolerant error budget, since Clifford gates,
measurements, idling, and magic-state-factory operations must ultimately share the allowed failure probability.

Determining whether this target can be reached with a particular
fault-tolerant architecture requires an additional layer of resource
estimation beyond the present work. A surface-code implementation, for
example, would require a specified physical error rate and noise model to
determine the code distance and logical error per code cycle~\cite{fowler2012surface}. The
$T$-state requirement would then need to be compared with a chosen
magic-state injection and distillation protocol to determine whether one or
multiple distillation rounds are required, together with the corresponding
factory footprint and throughput~\cite{bravyi2005universal,bravyi2012magic,o2017quantum}. A complete estimate would further include
logical Clifford and measurement errors, physical-qubit overhead, logical
clock speed, factory parallelism, and end-to-end runtime~\cite{litinski2019game,beverland2022assessing}. The present QCNO
analysis supplies the application-dependent inputs to that calculation:
the TEBD2 accuracy scale, the synthesis tolerance, the geometry-dependent
rotation workload, and the resulting logical $T$ count. Extending these
results to larger particle numbers and spatial domains and combining them
with a specific error-correcting architecture is therefore the next step
toward determining whether this neutrino simulation can provide a practical
early-fault-tolerant quantum-utility benchmark.

\section{Conclusions}
\label{sec:conclusions}

Neutrino flavor evolution in dense astrophysical environments is a challenging
MB problem because quantum correlations and spatial structure can drive
dynamics beyond MF descriptions, raising the question of whether the
suppression of FFI-like one-body coherence is a genuine correlated effect or a
finite-resource artifact. Because exact statevector evolution becomes exponentially expensive with \(N\) and tensor-network methods remain scalable only while the evolving state can be compressed efficiently, the goal is not simply to simulate more qubits, but to determine the numerical accuracy and quantum resources required to preserve the physics of interest. We
developed \texttt{QCNO} as a quantum simulation code for the
time-dependent inhomogeneous problem, including advection, position-dependent
interactions, and boundary conditions, using a common TEBD2 circuit across
ideal MB simulation, backend-aware execution, and logical
Clifford+$T$ resource estimation. This framework lets us address three
connected co-design questions within the same workflow: what system size and
numerical resolution are needed to preserve the MF--MB physics, whether that
signal survives backend compilation and hardware noise, and what logical
resources are required at the accuracy set by the application. Throughout these stages we demonstrate that spatial geometry is not only part of the neutrino physics but also has a huge advantage for the quantum simulation: it determines the active interaction graph, the hardware workload, and the accumulation of logical resources, connecting the physical question to the computational resources required for a scientifically useful quantum simulation.

We reproduce the MF--MB behavior reported in
Ref.~\cite{laraib2025two} and extend the exact MB comparison to the largest
system size $N=30$ for this application: The MF evolution develops the expected fast flavor
instability through exponential transverse Polarization growth, whereas the correlated MB
evolution does not. However, in MB simulations the longitudinal component is rapidly depolarized and its
finite-size oscillations decrease as $N$ increases. 
Its corresponding state complexity through R\'enyi-entropy and
stabilizer-R\'enyi-magic diagnostics show that this loss of one-body
polarization is accompanied by the growth of entanglement and non-stabilizer
structure (as also found in \cite{froustey2026collective,chernyshev2025quantum}). A single open-boundary crossing is sufficient to generate this
many-body complexity, after which the diagnostics approach plateaus, while
continued interactions in the closed geometry sustain its evolution. A production timestep of
$dt=5\times10^{-13}\ {\rm s}=0.047\mu^{-1}$ preserves the polarization-level
conclusions without adding unnecessary computational cost.

Favorable circuit scaling does not by itself
guarantee accurate NISQ execution. The localized interaction structure in our problem gives
approximately linear gate growth with the number of qubits $N$ across the backends studied, but
noisy executions still produce
$\epsilon({\langle P_z\rangle})_{\rm RMSE}\sim0.1$--$0.3$, while the ideal
IonQ simulator remains close to the Aer reference. At
$N_{\rm shots}=1024$, finite sampling is already subdominant; additional shots
reduce statistical uncertainty but not the backend-dependent damping, bias,
or routing- and gate-induced errors. The mitigation strategies tested here
also provide only marginal or inconsistent improvement. The main near-term
limitation is therefore whether connectivity, routing, native-gate noise,
and accumulated depth preserve the required MB signal.

This motivates the transition to error-corrected execution, where logical
errors can be suppressed with increasing code distance, provided the physical
error rate is below threshold, so that hardware noise can be reduced below
the application-level numerical error budget~\cite{fowler2012surface}. Fault-tolerant co-design asks how much logical accuracy is required
by the application. For our canonical circuits, the accumulated Trotter
error is of order
$\epsilon_{\rm TEBD2}\sim10^{-3}$, requiring a per-rotation synthesis
tolerance of $\epsilon_{\rm syn}\sim10^{-7}$, for which each retained $R_Z$
requires approximately $70$ $T$ gates. Tightening the synthesis tolerance beyond this
scale would add non-Clifford cost without improving the resolved physics
unless the TEBD2 error were also reduced. We report the scaling of the resources required in Equations~\ref{eq:rz_estimated_t_count_methods}, \ref{eq:ntrotter_estimate}, and \ref{eq:esyn_final_estimate}. For the open-boundary case, the time-dependent interaction graph produces
nonlinear $T$-count accumulation across Trotter substeps, reducing the cost by a factor of 2 over the time when the two beams interact. For this circuit, an error-corrected quantum computer would need a logical-gate accuracy of \(10^{-7}\) to support the \(\mathcal{O}(10^4)\)-\(T\)-gate neutrino circuit while keeping logical faults at the \(10^{-3}\)(about $99.9\%$ probability of completing the circuit without
such a fault), assuming independent logical $T$-gate errors dominate the
failure budget.

These results identify both the opportunity and the remaining
gap toward quantum utility for simulation of many-body neutron flavor transformation physics. We do not establish the
asymptotic large-$N$ behavior of the problem, quantum advantage, or a universal
NISQ-to-fault-tolerant crossover. The detailed entanglement, magic, and
fault-tolerant resource calculations remain restricted to smaller systems,
the compiled gate exponents are finite-size and compiler dependent, and the
two-flavor two-beam model does not yet include realistic energy and angular
distributions, matter effects, collisions, three-flavor evolution, or
multidimensional transport. A complete fault-tolerant estimate must also
specify the physical error model, error-correcting code, code distance,
logical failure allocation, magic-state injection and distillation,
physical-qubit overhead, factory throughput, and total runtime~\cite{beverland2022assessing,o2017quantum}. 

Instead, the present analysis identifies the critical, application-dependent inputs to that crossover: the timestep required to resolve the dynamics, the interaction history determining the rotation workload, and the synthesis accuracy required relative to the Trotter error. Future work
should therefore extend the MB calculations to larger $N$, longer
beams, and larger domains; generalize the circuit to the full
momentum-dependent Hamiltonian ~\cite{cirigliano2024neutrino} or use a semi-classical approach to the non-forward processes in full hamiltonian~\cite{carlson2026neutrino}; introduce more realistic neutrino distribution such as wave-packet treatment~\cite{cervia2026interactions};  investigate structure-exploiting Hamiltonian-approximation
methods, including fast-forwarding where applicable~\cite{berry2007efficient,low2019hamiltonian,
atia2017fast}; and perform broader
matched live-hardware and error-mitigation studies before connecting the
resulting application targets to a complete logical architecture.

\section{Acknowledgments}
We would like to thank Antigoni Georgiadou for useful discussions. SR and ZL are supported by NSF Grant no. PHY-2412683. This research used resources of the Oak Ridge Leadership Computing Facility at Oak Ridge National Laboratory, which is supported by the Office of Science of the US Department of Energy under Contract No. DE-AC05-00OR22725.
This manuscript has been authored in part by UT-Battelle, LLC under Contract No. DE-AC05-00OR22725 with the U.S. Department of Energy. The United States Government retains and the publisher, by accepting the article for publication, acknowledges that the United States Government retains a non-exclusive, paid-up, irrevocable, worldwide license to publish or reproduce the published form of this manuscript, or allow others to do so, for United States Government purposes. The Department of Energy will provide public access to these results of federally sponsored research in accordance with the DOE Public Access Plan. (http://energy.gov/downloads/doe-public-279 access-plan).

\appendix
\section{QCNO physics validation tests}
\label{app:qcno_validation}

Before applying QCNO to the spatially inhomogeneous system studied in the main text, we
validate the implementation against known MF and MB neutrino calculations and their analytical limits. These tests separately check the vacuum
Hamiltonian, the neutrino self-interaction Hamiltonian, and their combined Homogeneous evolution.The purpose of these benchmarks is not to reproduce the full analysis of
Refs.~\cite{roggero2021entanglement,richers2021particle}, but to show that the same QCNO
implementation used for the new calculations recovers established limits and published
neutrino-oscillation results. We use
Qiskit Aer direct state and count-based during backend execution layer (See Fig~\ref{fig:qcno_code_framework}) results in this appendix section. Backend effects are
tested for the inhomogeneous application in the main text and subsequent Appendix section ~\ref{section:backend_open_closed_appendix}. We make heavy use of Qiskit \cite{javadiabhari2024quantumcomputingqiskit}.

\subsection{Vacuum oscillations}
\label{sec:classical_vac_osc}

\begin{figure}
\centering
\includegraphics[width=\linewidth]{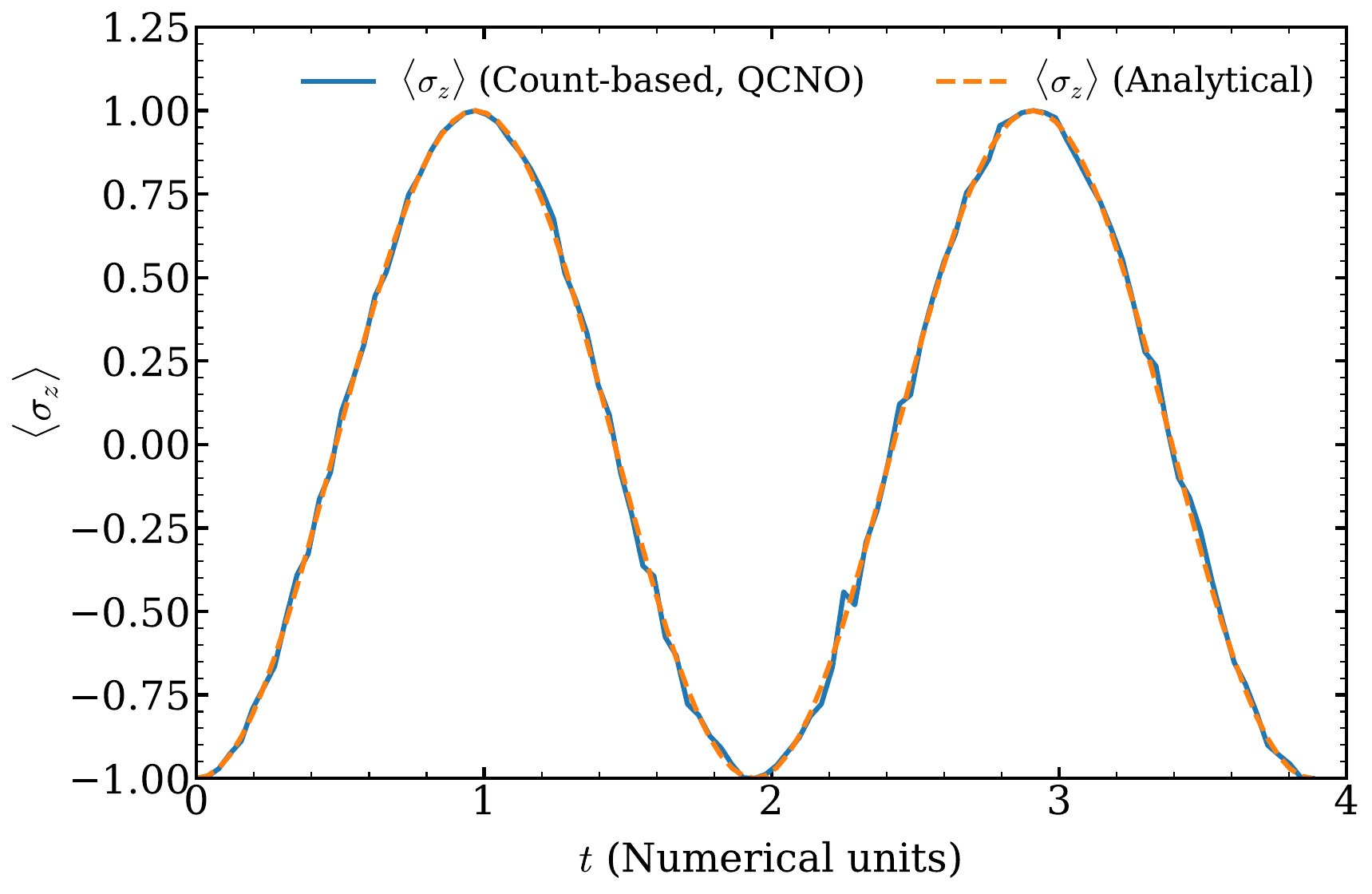}
\caption{
Vacuum-oscillation validation for $N = 4$ for 1024 shots with $\mu = 0$(scaled appropriately on x-axis),
$\omega_i = \pi~\mathrm{s}^{-1}$, and $\vec{B} = (1,0,0)$. The QCNO count-based Qiskit Aer calculation
reproduces the expected sinusoidal evolution of the longitudinal polarization,
$\langle \sigma^z_i(t)\rangle = -\cos(\omega_i t)$, for site $i = 1$.
}
\label{fig:vac_osc_validation}
\end{figure}
We first test the one-body vacuum Hamiltonian by setting the self-interaction strength to
$\mu = 0$. We use $N = 4$ particles divided equally
between two initially pure flavor populations, beam $A$ ($i \in [1,2]$) and beam $B$
($i \in [3,4]$), with $\omega_{i \in A} = \omega_A$ and $\omega_{i \in B} = \omega_B$. Every
particle is given $
    \omega_i = \pi~\mathrm{s}^{-1},
    \vec{B} = (1,0,0),
$ which we set through $\Delta m^2 = 2\pi~\mathrm{s}^{-2}$, $E_i = 1$, and $s_i = 1$ (all
neutrinos, no antineutrinos). The initial state is the product state
$
    |\Psi_0\rangle
    = \bigotimes_{n \in A}^{N/2} |\!\downarrow\rangle_n
    \otimes
    \bigotimes_{m \in B}^{N/2} |\!\uparrow\rangle_m ,
$
so the Hamiltonian governing the evolution using Eq.~\ref{eq:qcno_vacuum} becomes
\begin{equation}
    H_\mathrm{vac} = \frac{\omega_A}{2} \sum_{i \in A} \sigma_i^x
                   + \frac{\omega_B}{2} \sum_{i \in B} \sigma_i^x \, .
    \label{eq:Hvac_AB}
\end{equation}
With these choices each particle rotates in the $y$--$z$ plane, and the expected solution for
a particle initialized in $|\!\downarrow\rangle$ is
$
    \langle \sigma_i^z(t)\rangle = -\cos(\omega_i t).
$
Figure~\ref{fig:vac_osc_validation} shows that the QCNO Qiskit Aer evolution reproduces this
analytic oscillation for site $i = 1$. This test checks the vacuum term and the corresponding
single-qubit rotations used in the circuit construction.

\subsection{Many-body neutrino self-interactions}
\label{sec:self_int_validation}
\begin{figure}
\centering
\includegraphics[width=0.95\linewidth]{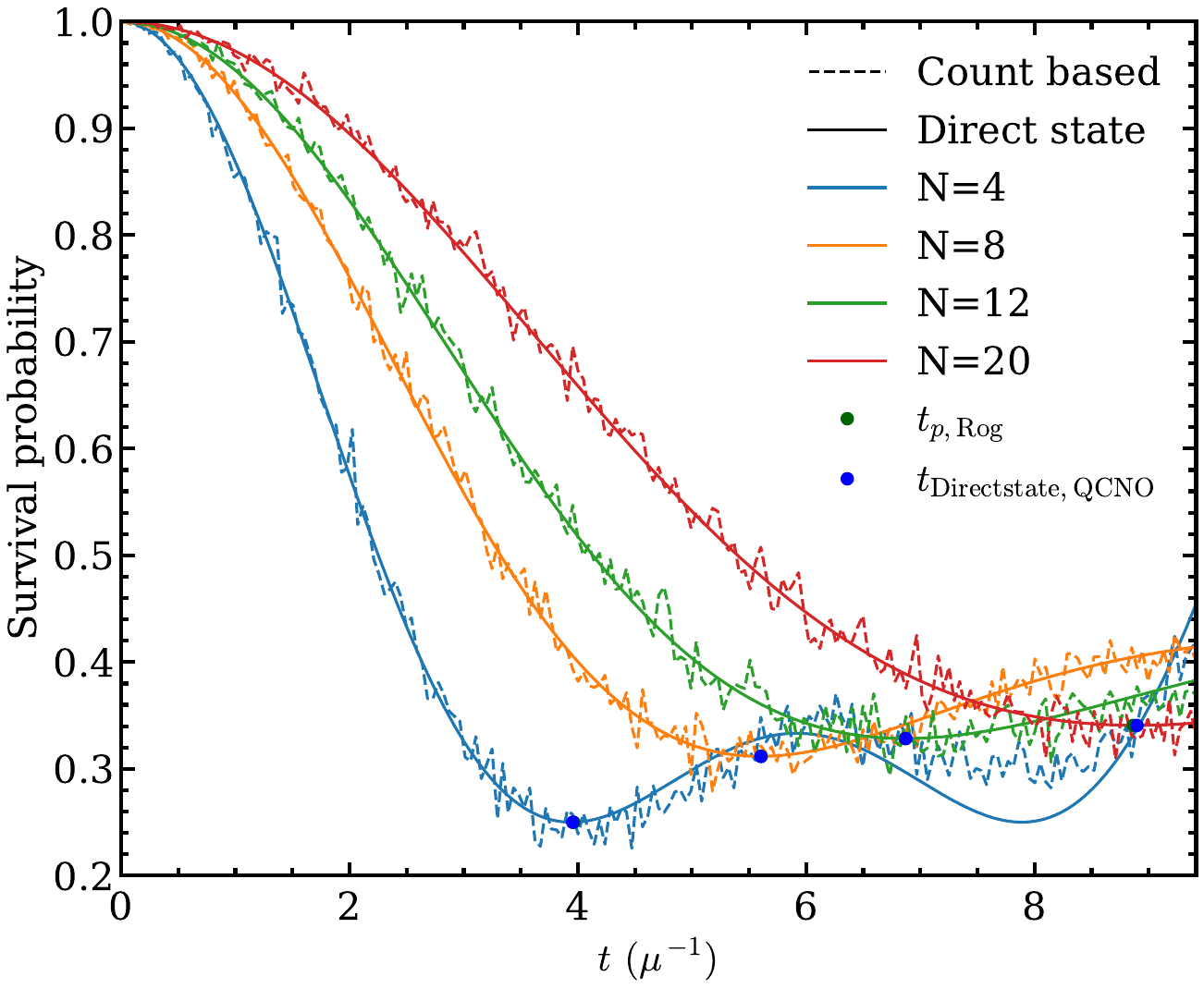}
\caption{
MB self-interaction validation for $N = 4, 8, 12,$ and $20$ with
$\omega = 0$. QCNO Qsikit Aer direct state(solid) and count-based(dashed)results are compared for 1024 shots. Their survival-probability trajectories agree within this shot size for all system sizes tested (in different colors) along with our first local minima agreement with the $t_p$ in
Ref.~\cite{roggero2021entanglement}. 
}
\label{fig:self_int_validation}
\end{figure}

We next test the MB self-interaction Hamiltonian by setting the vacuum frequency to
$\omega = 0$ and using the homogeneous and isotropic limit of Eq.~(\ref{eq:qcno_self_interaction}). Each
numerical site N holds
$\mathcal{N}= \sum_{i=1}^{N} \mu\,\Delta x^3 / (\sqrt{2}\,G_F N)$
physical neutrinos with $\mu$ initialized to unity, and makes the dynamics depend only on the scaled time $\mu t$.
The particles are initialized in the product state
$
    |\Psi_0\rangle
    = \bigotimes_{i=1}^{N/2} |\!\downarrow\rangle_i
    \otimes
    \bigotimes_{i=N/2+1}^{N} |\!\uparrow\rangle_i ,
$
with equal populations in the two initial flavor states. We calculate the single-particle
survival probability $\rho_{ee,i} $  (Eq.~(\ref{eq:reduced_density_II})
for $N = 4, 8, 12,$ and $20$. Figure~\ref{fig:self_int_validation} compares the
QCNO Direct-state and count-based Qiskit Aer circuit MB evolution. The
trajectories agree over the tested system sizes for the 1024 shots considered. Increasing shots will further remove this statistical uncertainty between the solid and dashed lines by $\sqrt{N_\mathrm{shots}}$. Our results also reproduce the location of the first
survival-probability local minimum $t_p$ (green scatter point) in units of $\mu^{-1}$ (reported in Fig.~3(b) of Roggero et
al.~\cite{roggero2021entanglement}) which is hidden exactly underneath the QCNO local minima (blue scatter point). These minima are also consistent
across system-size dependence. This test checks the MB pair-interaction term and
its implementation as the number of interacting particles is increased.

\subsection{Homogeneous mean-field fast flavor instability}
\label{sec:homo_ffi_validation}
\begin{figure}
\centering
\includegraphics[width=\linewidth]{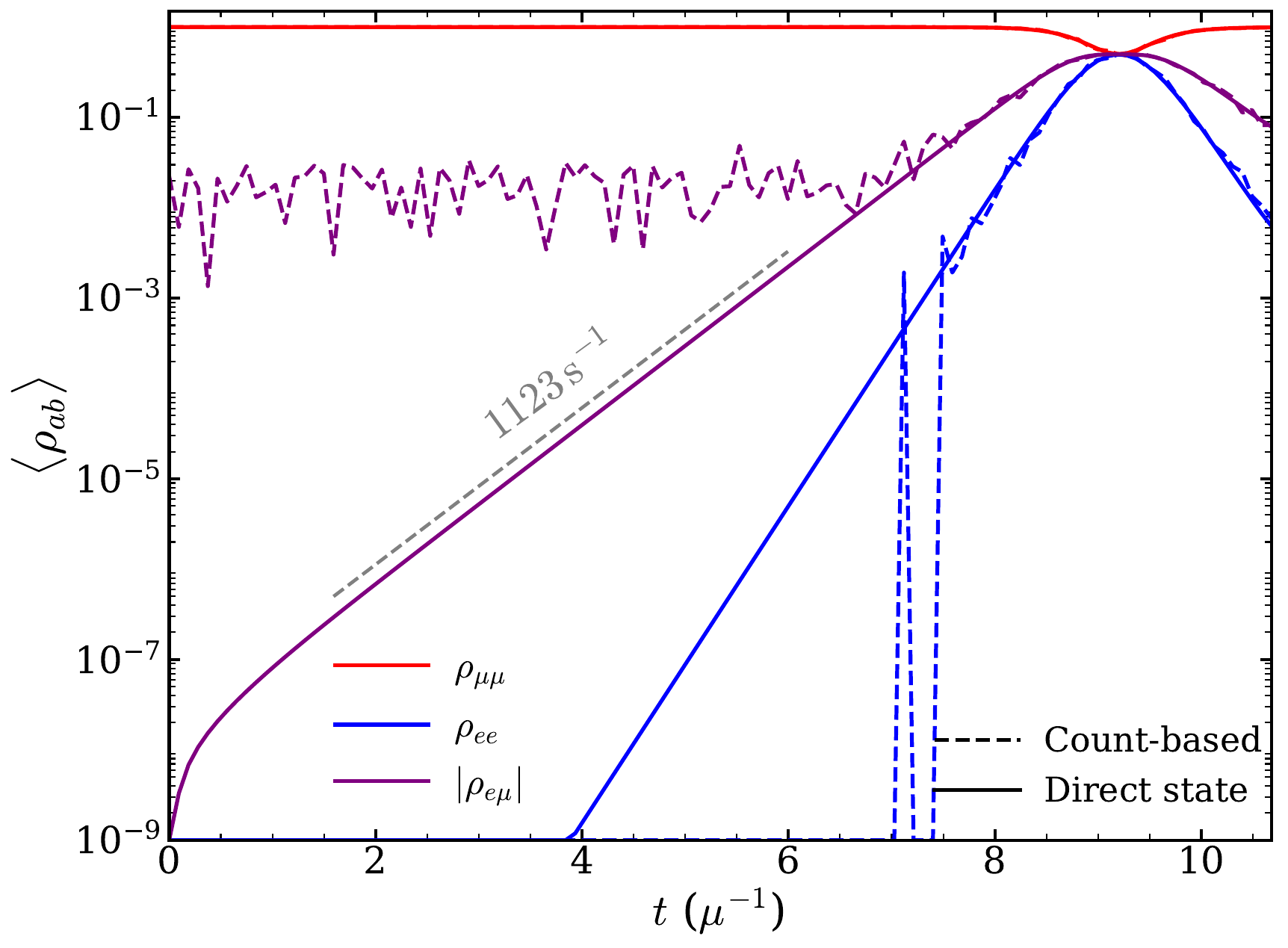}
\caption{
Homogeneous MF fast-flavor-instability validation using the setup of
Ref.~\cite{richers2021particle}. The direct QCNO calculation (solid) reproduces exponential
growth of the off-diagonal flavor coherence at the linear-stability rate
$1123~\mathrm{s^{-1}}$, followed by nonlinear flavor conversion. The shot-based result
(dashed) agrees once $|\rho_{e\mu}|$ exceeds the statistical noise floor near $t = 12.45$~ms $\approx 7 \, \mu^{-1}$.
}
\label{fig:homo_ffi_validation}
\end{figure}

Finally, we test the combined Hamiltonian in a case with a known MF fast flavor
instability. We use the homogeneous two-beam setup of Ref.~\cite{richers2021particle}: two
oppositely moving particles in domain with $\Delta x = 10^7$~cm, one electron
neutrino moving in $-\hat{x}$ and one electron antineutrino moving in $+\hat{x}$, both with
$E_i = |p_i| = 50$~MeV. The masses follow the inverted hierarchy,
$
    m_2 = 0.008596511~\mathrm{eV},
    m_1 = 0,
$
with mixing angle
$
    \theta_\nu = \theta_{12} = 1.74532925\times10^{-8},
$
so the vacuum Bloch vector in the flavor basis is
$\vec{B} = [-\sin 2\theta_\nu,\, 0,\, \cos 2\theta_\nu]$ and
$\omega_i = \Delta m^2 / 2E_i$. The number densities are
$n = (m_2^2 - m_1^2)c^4 / (4\sqrt{2}\,G_F h \nu) = 2.92\times10^{24}~\mathrm{cm^{-3}}$, which
sets the interaction strength $\mu = \sqrt{2}\,G_F n= 1.77856\times10^{-3}$ s). The interaction is homogeneous,
$S(\xi_{ij}) = 1$, and fixing the bond dimension to one restricts the evolution to the
factorized MF limit of Eq.~(\ref{eq:qcno_full_hamiltonian} where $\langle P_i \rangle = \langle \sigma_i \rangle $ for i-th site). We track the diagonal flavor
populations through the density matrix components $\rho_{ee} $
$\rho_{\mu\mu}$ and the off-diagonal coherence
$ |\rho_{e\mu}| $ (Eq.~(\ref{eq:reduced_density_II})
The direct state QCNO calculation in Fig.~\ref{fig:homo_ffi_validation} reproduces the expected
exponential growth of the off-diagonal component followed by nonlinear flavor conversion,
matching Fig.~14 of Ref.~\cite{richers2021particle}. The fitted linear growth rate agrees
with the analytic linear-stability result
$\mathrm{Im}(\Omega) = (m_2^2 - m_1^2)c^4/(2\hslash h \nu) = 1123~\mathrm{s^{-1}}$. However, resolving
this rate requires direct state measurement of the density matrix, since a shot-based measurement
sits on a noise floor of order $N_\mathrm{shots}^{-1/2}$ and cannot see $|\rho_{e\mu}|$ until
it has already grown above that floor. We show the count-based (dashed) density matrix components with 1024 shots for reference. This test shows that QCNO reproduces the known
collective Fast Flavor instability when the vacuum and self-interaction terms are used together.

\section{Inhomogeneous FFI}
\subsection{TEBD2 Product-formula timestep convergence}
\label{sec:appendix:Numerical_precision}

\begin{figure}
\centering
\includegraphics[width=1.0\linewidth]
{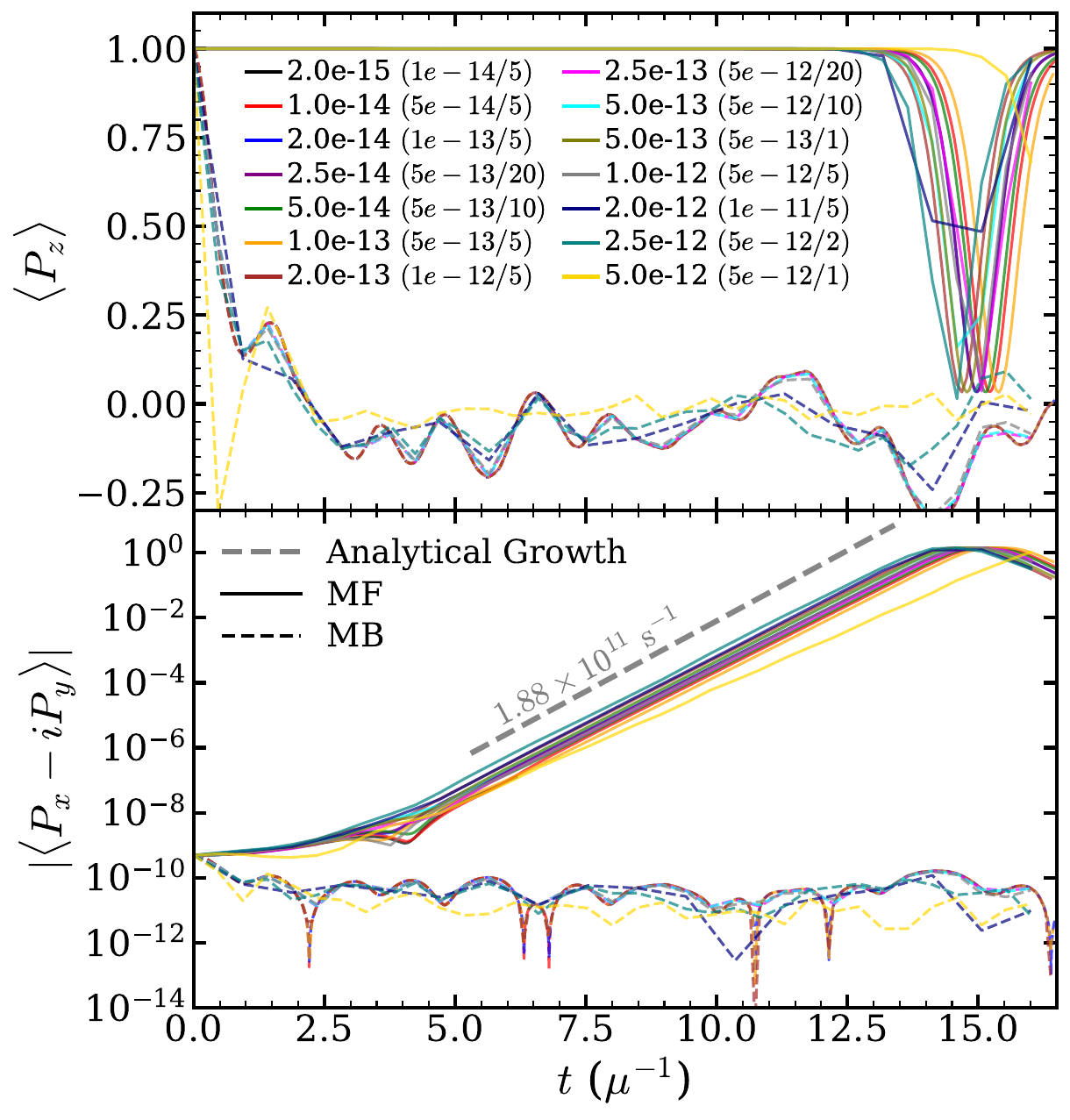}
\caption{
TEBD2 Product-formula timestep dependence of the $N=14$ inhomogeneous two-beam
system. The upper panel shows the longitudinal polarization
$\langle P_z\rangle$, and the lower panel shows the transverse coherence
$\left|\langle P_x-iP_y\rangle\right|$. Solid and dashed curves show MF and
MB evolution, respectively, for different values of
$dt=\mathrm{t_{out}}/N_{\rm steps}$. The gray dashed line shows the analytic
MF growth rate, $1.88\times10^{11}\ {\rm s}^{-1}$. Large timesteps shift the
MF growth and nonlinear flavor conversion and also change the early MB
evolution. The results become stable for
$dt\lesssim5\times10^{-13}\ {\rm s}=0.047\mu^{-1}$, which is used for the
production calculations. 
}
\label{fig:precision_dt_all}
\end{figure}

Figure~\ref{fig:precision_dt_all} tests timestep convergence for the
$N=14$ inhomogeneous two-beam system using direct Qiskit Aer simulation
and the same second-order product formula used in the production runs.
We vary $\mathrm{t_{out}}$ and $N_{\rm steps}$, with
$dt=\mathrm{t_{out}}/N_{\rm steps}$, over
$2\times10^{-15}\ {\rm s}\leq dt\leq 5\times10^{-12}\ {\rm s}$. Large timesteps
shift the MF transverse growth and the time at which $\langle P_z\rangle$
changes, and also distort the early MB depolarization. These differences
decrease as $dt$ is reduced. For
$dt\lesssim5\times10^{-13}\ {\rm s}=0.047\mu^{-1}$, both MF and MB
polarization trajectories are stable. The MF calculation still shows
exponential transverse growth with the expected linear-stability rate
$1.88\times10^{11}\ {\rm s}^{-1}$
~\cite{chakraborty2016self2,richers2021particle}, followed by nonlinear
flavor conversion, while the MB calculation shows rapid loss of
longitudinal polarization without the same growing transverse mode. The
MF--MB difference in the main text is therefore not caused by the
product-formula timestep. Two runs with the same
$dt=5\times10^{-13}\ {\rm s}$, using
$(\mathrm{t_{out}},N_{\rm steps})=(5\times10^{-12}\ {\rm s},10)$ and
$(5\times10^{-13}\ {\rm s},1)$, also agree. We therefore use
$\mathrm{t_{out}}=5\times10^{-13}\ {\rm s}$,
$N_{\rm steps}=1$, and $dt=5\times10^{-13}\ {\rm s}$ for the production
calculations. This test shows convergence of the polarization observables
used here, not of the full MB wavefunction. Smaller timesteps do not change
the physical conclusions and would only increase the number of
product-formula layers and the circuit cost.

\subsection{Backend-based evolution under different geometries}
\label{section:backend_open_closed_appendix}
We compare the MB evolution across several backend models introduced in
Sec.~\ref{sec:quantum_framework}. These include the IBM backend Guadalupe
(16 available qubits, simulated) and \texttt{ibm\_boston}
(156 available qubits, hardware), together with the ideal IonQ cloud
simulator (25 available qubits) and the IonQ cloud simulator using the
\texttt{aria-1} noise model (25 available qubits). Qiskit Aer is used as the
ideal reference. The backend connectivity, native gates, and noise properties
are described in Sec.~\ref{sec:hardware_backend_sensitivity}. For each
backend, we show results up to the largest system size supported by its
available qubit count and computational cost. Since the sweeps shown here
start at $N=6$, \texttt{FakeManilaV2} does not appear in these figures.

\begin{figure*}
    \centering
    \includegraphics[width=\textwidth]{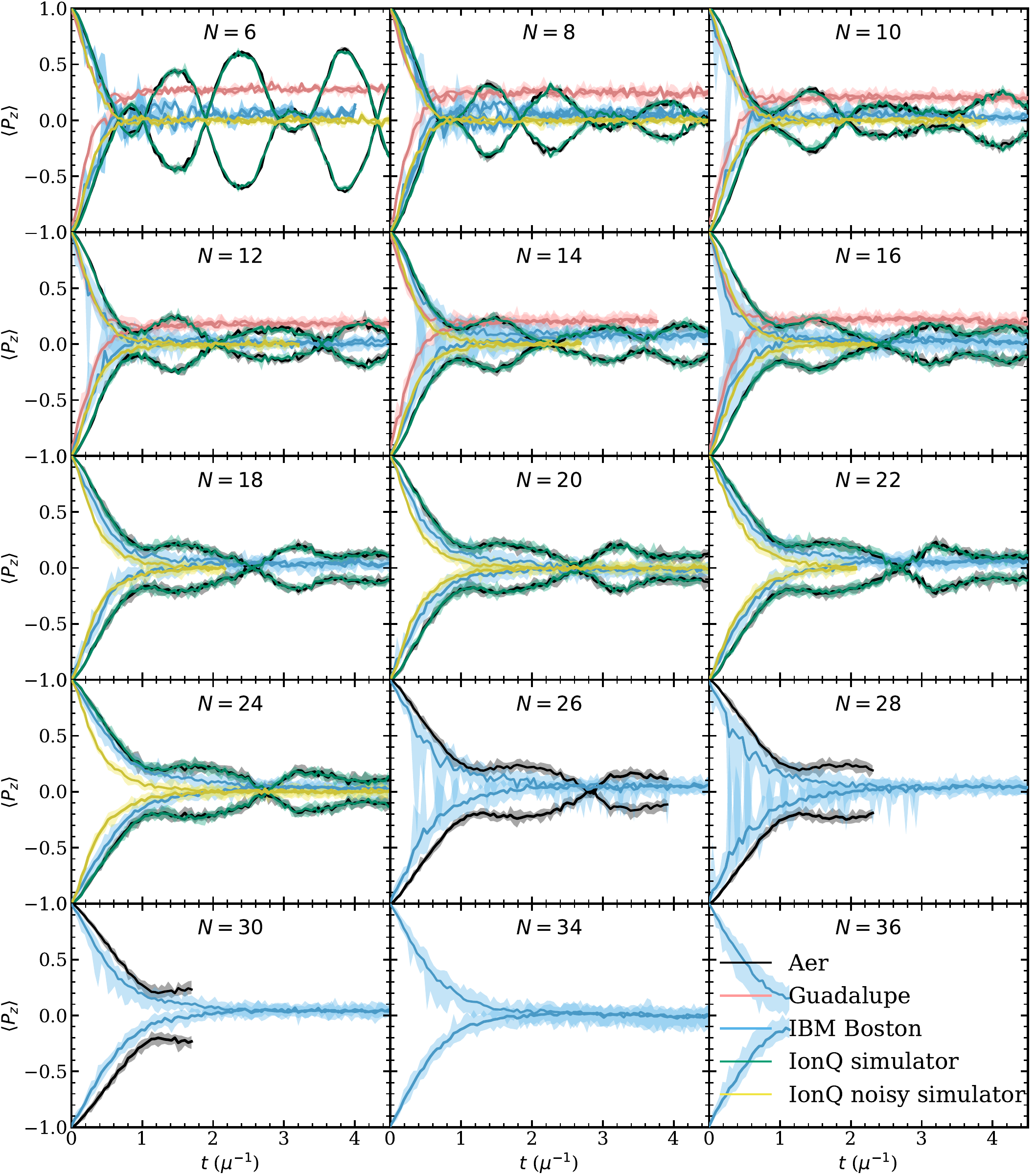}
    \caption{
    Backend dependence of the closed-boundary MB longitudinal polarization
    across system size. Each panel fixes $N$ and shows the two initial flavor
    beams. Lines give the mean $\langle P_z\rangle$ over each beam half and
    shaded regions span the site-wise minimum to maximum. Colors denote
    Qiskit Aer, the IBM Guadalupe backend model, IBM Boston, the ideal IonQ
    simulator, and the IonQ noisy simulator. All calculations use the same
    physical model and product-formula timestep
    $d t=5\times10^{-13}~{\rm s}=0.047\mu^{-1}$ and  1024 shots. Aer provides the ideal reference.
    The ideal IonQ simulator closely follows the Aer dynamics, while noisy
    implementations progressively damp or bias the finite-size oscillatory
    structure.
    }
    \label{fig:hardware_backend_closed}
\end{figure*}
Figure~\ref{fig:hardware_backend_closed} shows the backend comparison for
closed boundaries upto 36 qubits. The Aer and ideal IonQ results show how behavior of flavor equilibration ($\langle P_z\rangle\rightarrow0$) changes with system size without gate noise. At small $N$, the
system does not remain near flavor equilibrium after the initial
depolarization. Instead, the closed boundaries cause the particles to
repeatedly interact as they move through the periodic domain, producing large
oscillations in $\langle P_z\rangle$ over time. As $N$ increases, these
finite-size oscillations become smaller and the beam-averaged polarization
stays closer to zero, so the system reaches and remains near flavor
equilibration more clearly. The shaded regions show the minimum-to-maximum
spread across sites within each beam and also become smaller as the large
finite-size oscillations are reduced. This system-size trend is seen in both
Aer and the ideal IonQ simulator and gives the noiseless reference for the
physical evolution. The small fluctuations in the ideal IonQ results come
from finite-shot sampling($\sigma_{\langle P_z\rangle}\propto N_{\rm shots}^{-1/2}$). The other backends contain additional effects that
change this ideal behavior. Gate noise in the IonQ noisy simulator damps the
oscillations and can drive $\langle P_z\rangle$ toward zero even when the
corresponding ideal trajectory still contains structure. Guadalupe shows a
different error pattern, with the trajectories developing a persistent
positive offset of about $\langle P_z\rangle\sim0.2$ across the available
system sizes rather than remaining centered around zero, as also discussed in
Sec.~\ref{sec:hardware_backend_sensitivity}. The IBM backends also have
restricted coupling maps, so the nonlocal interactions require extra routing,
which adds two-qubit gates and circuit depth. However, IBM Boston follows the
ideal evolution better than Guadalupe even though both have restricted
connectivity. This is consistent with the lower gate errors of the Boston
device and shows that connectivity alone does not determine the backend
performance. The physical system-size dependence of flavor equilibration is
therefore identified from Aer and the ideal IonQ simulator, while the other
backends show how finite-shot sampling, gate noise, and routing change that
ideal evolution.

\begin{figure*}
    \centering
    \includegraphics[width=0.9\textwidth]
    {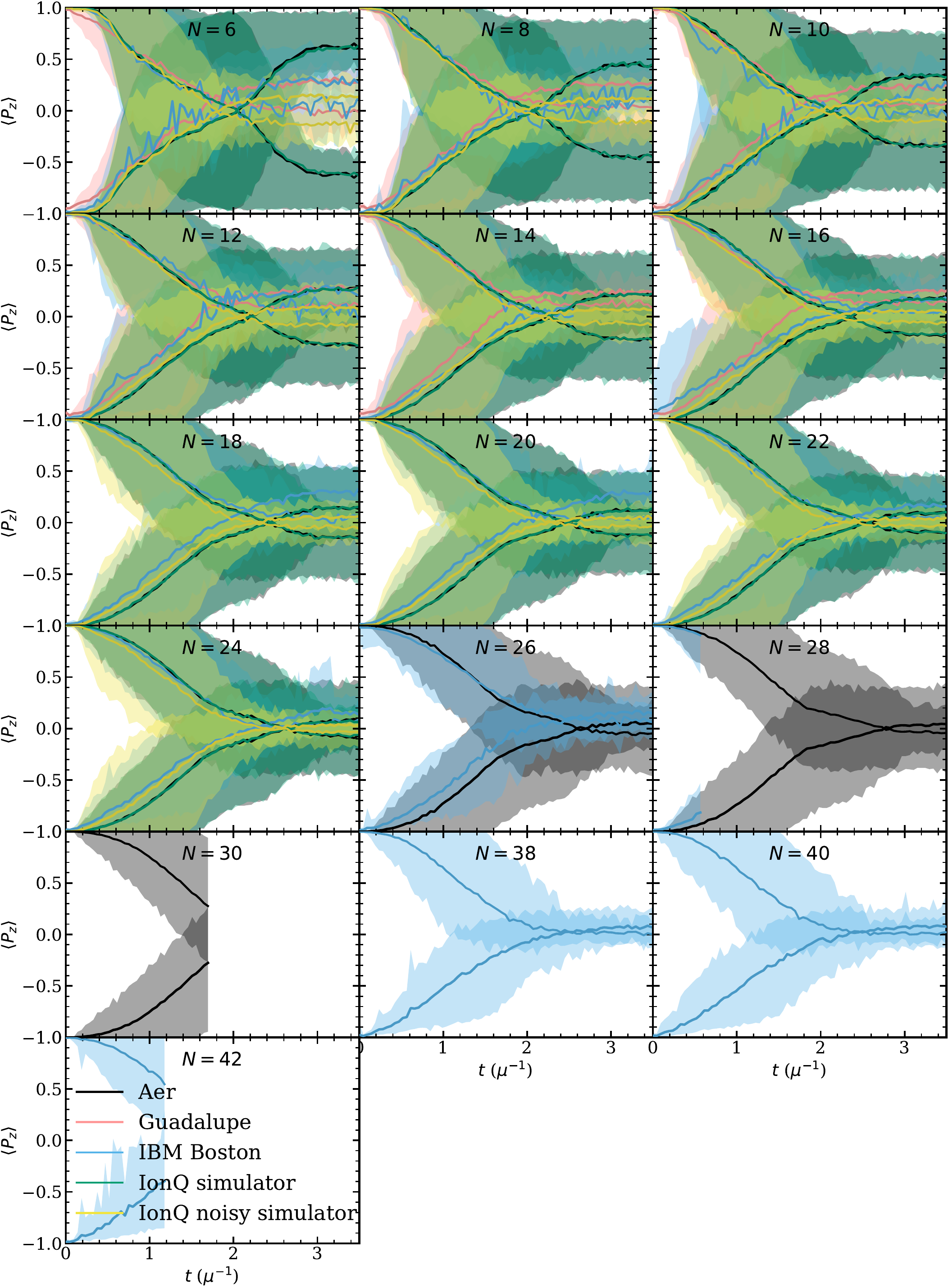}
    \caption{
    Similar to Fig.~\ref{fig:hardware_backend_closed} but for open boundaries where beams (from opposite ends) interact during a
    single crossing and then separate. The broad site-wise spead reflect the
    different spatial interaction histories of individual particles.
    }
    \label{fig:hardware_backend_open}
\end{figure*}
Figure~\ref{fig:hardware_backend_open} shows the corresponding backend
comparison for open boundaries. The main backend effects are the same as in
the closed case: the ideal IonQ simulator follows the Aer reference apart from
finite-shot fluctuations, the IonQ noisy simulator damps the signal, and
Guadalupe again shows the positive offset discussed above, while the IBM
backends are also affected by routing on their restricted coupling maps. The
main difference is the open-boundary geometry. Here, the two beams interact
only while they overlap, so the number of active interaction pairs changes
with time and drops as the beams separate. This reduces the amount of
interaction evolution that must be implemented and limits how long gate and
routing errors can accumulate. This is also consistent with the smaller RMSE
errors for the open geometry shown in
Fig.~\ref{fig:hardware_backend_summary}. Because of this shorter interaction
window and the reduced number of active pairs, the open-boundary calculations
remain practical to larger system sizes, and we show backend results up to
$N=42$. The Aer and ideal IonQ results again give the physical reference:
with increasing $N$, the beam-averaged $\langle P_z\rangle$ evolution becomes
smoother and approaches flavor equilibration during the crossing, while the
shaded minimum-to-maximum site spread remains broader than in the closed case
because different sites experience different overlap times. The noisy
backends then modify this ideal evolution during the finite crossing, but the
errors have less time to build up than for closed boundaries.

\bibliography{apssamp}

@article{Childs2021Trotter,
  title   = {Theory of Trotter Error with Commutator Scaling},
  author  = {Childs, Andrew M. and Su, Yuan and Tran, Minh C. and
             Wiebe, Nathan and Zhu, Shuchen},
  journal = {Physical Review X},
  volume  = {11},
  pages   = {011020},
  year    = {2021},
  doi     = {10.1103/PhysRevX.11.011020}
}

@article{turro2025qutrit,
  title={Qutrit and qubit circuits for three-flavor collective neutrino oscillations},
  author={Turro, Francesco and Chernyshev, Ivan A and Bhaskar, Ramya and Illa, Marc},
  journal={Physical Review D},
  volume={111},
  number={4},
  pages={043038},
  year={2025},
  publisher={APS}
}

@misc{javadiabhari2024quantumcomputingqiskit,
      title={Quantum computing with Qiskit}, 
      author={Ali Javadi-Abhari and Matthew Treinish and Kevin Krsulich and Christopher J. Wood and Jake Lishman and Julien Gacon and Simon Martiel and Paul D. Nation and Lev S. Bishop and Andrew W. Cross and Blake R. Johnson and Jay M. Gambetta},
      year={2024},
      eprint={2405.08810},
      archivePrefix={arXiv},
      primaryClass={quant-ph},
      url={https://arxiv.org/abs/2405.08810}, 
}

@article{RossSelinger2016,
  title   = {Optimal ancilla-free Clifford+T approximation of z-rotations},
  author  = {Ross, Neil J. and Selinger, Peter},
  journal = {Quantum Information and Computation},
  volume  = {16},
  number  = {11--12},
  pages   = {901--953},
  year    = {2016},
  eprint  = {1403.2975},
  archivePrefix = {arXiv}
}

@article{pantaleone1992neutrino,
  title={Neutrino oscillations at high densities},
  author={Pantaleone, James},
  journal={Physics Letters B},
  volume={287},
  number={1-3},
  pages={128--132},
  year={1992},
  publisher={Elsevier}
}

@article{duan2010collective,
  title={Collective neutrino oscillations},
  author={Duan, Huaiyu and Fuller, George M and Qian, Yong-Zhong},
  journal={Annual Review of Nuclear and Particle Science},
  volume={60},
  pages={569--594},
  year={2010},
  publisher={Annual Reviews}
}

@article{sawyer2009multiangle,
  title={Multiangle instability in dense neutrino systems},
  author={Sawyer, Raymond F},
  journal={Physical Review D},
  volume={79},
  number={10},
  pages={105003},
  year={2009},
  publisher={APS}
}

@article{sawyer2005speed,
  title={Speed-up of neutrino transformations in a supernova environment},
  author={Sawyer, Raymond F},
  journal={Physical Review D},
  volume={72},
  number={4},
  pages={045003},
  year={2005},
  publisher={APS}
}

@article{sigl1993general,
  title={General kinetic description of relativistic mixed neutrinos},
  author={Sigl, G{\"u}nther and Raffelt, G},
  journal={Nuclear Physics B},
  volume={406},
  number={1-2},
  pages={423--451},
  year={1993},
  publisher={Elsevier}
}

@article{tamborra2021new,
  title={New developments in flavor evolution of a dense neutrino gas},
  author={Tamborra, Irene and Shalgar, Shashank},
  journal={Annual Review of Nuclear and Particle Science},
  volume={71},
  pages={165--188},
  year={2021},
  publisher={Annual Reviews}
}

@article{mckellar1994oscillating,
  title={Oscillating neutrinos in the early Universe},
  author={McKellar, Bruce HJ and Thomson, Mark J},
  journal={Physical Review D},
  volume={49},
  number={6},
  pages={2710},
  year={1994},
  publisher={APS}
}

@article{richers2021particle,
  title={Particle-in-cell simulation of the neutrino fast flavor instability},
  author={Richers, Sherwood and Willcox, Don E and Ford, Nicole M and Myers, Andrew},
  journal={Physical Review D},
  volume={103},
  number={8},
  pages={083013},
  year={2021},
  publisher={APS}
}

@article{capozzi2022neutrino,
  title={Neutrino flavor conversions in high-density astrophysical and cosmological environments},
  author={Capozzi, Francesco and Saviano, Ninetta},
  journal={Universe},
  volume={8},
  number={2},
  pages={94},
  year={2022},
  publisher={MDPI}
}

@article{vidal2003efficient,
  title={Efficient classical simulation of slightly entangled quantum computations},
  author={Vidal, Guifr{\'e}},
  journal={Physical review letters},
  volume={91},
  number={14},
  pages={147902},
  year={2003},
  publisher={APS}
}

@article{feynman1982simulating,
  title={Simulating physics with computers},
  author={Feynman, Richard P and others},
  journal={Feynman and computation},
  volume={22},
  pages={133--153},
  year={2018}
}

@article{cowtan2019qubit,
  title={On the qubit routing problem},
  author={Cowtan, Alexander and Dilkes, Silas and Duncan, Ross and Krajenbrink, Alexandre and Simmons, Will and Sivarajah, Seyon},
  journal={arXiv preprint arXiv:1902.08091},
  year={2019}
}

@article{huggins2021efficient,
  title={Efficient and noise resilient measurements for quantum chemistry on near-term quantum computers},
  author={Huggins, William J and McClean, Jarrod R and Rubin, Nicholas C and Jiang, Zhang and Wiebe, Nathan and Whaley, K Birgitta and Babbush, Ryan},
  journal={npj Quantum Information},
  volume={7},
  number={1},
  pages={23},
  year={2021},
  publisher={Nature Publishing Group UK London}
}

@article{bravyi2012magic,
  title={Magic-state distillation with low overhead},
  author={Bravyi, Sergey and Haah, Jeongwan},
  journal={Physical Review A—Atomic, Molecular, and Optical Physics},
  volume={86},
  number={5},
  pages={052329},
  year={2012},
  publisher={APS}
}

@article{litinski2019game,
  title={A game of surface codes: Large-scale quantum computing with lattice surgery},
  author={Litinski, Daniel},
  journal={Quantum},
  volume={3},
  pages={128},
  year={2019},
  publisher={Verein zur F{\"o}rderung des Open Access Publizierens in den Quantenwissenschaften}
}

@article{beverland2022assessing,
  title={Assessing requirements to scale to practical quantum advantage},
  author={Beverland, Michael E and Murali, Prakash and Troyer, Matthias and Svore, Krysta M and Hoefler, Torsten and Kliuchnikov, Vadym and Low, Guang Hao and Soeken, Mathias and Sundaram, Aarthi and Vaschillo, Alexander},
  journal={arXiv preprint arXiv:2211.07629},
  year={2022}
}

@article{fowler2012surface,
  title={Surface codes: Towards practical large-scale quantum computation},
  author={Fowler, Austin G and Mariantoni, Matteo and Martinis, John M and Cleland, Andrew N},
  journal={Physical Review A—Atomic, Molecular, and Optical Physics},
  volume={86},
  number={3},
  pages={032324},
  year={2012},
  publisher={APS}
}

@article{babbush2018encoding,
  title={Encoding electronic spectra in quantum circuits with linear T complexity},
  author={Babbush, Ryan and Gidney, Craig and Berry, Dominic W and Wiebe, Nathan and McClean, Jarrod and Paler, Alexandru and Fowler, Austin and Neven, Hartmut},
  journal={Physical Review X},
  volume={8},
  number={4},
  pages={041015},
  year={2018},
  publisher={APS}
}

@article{schollwock2011density,
  title={The density-matrix renormalization group in the age of matrix product states},
  author={Schollw{\"o}ck, Ulrich},
  journal={Annals of physics},
  volume={326},
  number={1},
  pages={96--192},
  year={2011},
  publisher={Elsevier}
}

@article{vidal2004efficient,
  title={Efficient simulation of one-dimensional quantum many-body systems},
  author={Vidal, Guifr{\'e}},
  journal={Physical review letters},
  volume={93},
  number={4},
  pages={040502},
  year={2004},
  publisher={APS}
}

@article{vidal2007classical,
  title={Classical simulation of infinite-size quantum lattice systems in one spatial dimension},
  author={Vidal, Guifr{\'e}},
  journal={Physical review letters},
  volume={98},
  number={7},
  pages={070201},
  year={2007},
  publisher={APS}
}

@article{verstraete2008matrix,
  title={Matrix product states, projected entangled pair states, and variational renormalization group methods for quantum spin systems},
  author={Verstraete, Frank and Murg, Valentin and Cirac, J Ignacio},
  journal={Advances in physics},
  volume={57},
  number={2},
  pages={143--224},
  year={2008},
  publisher={Taylor \& Francis}
}

@article{orus2014practical,
  title={A practical introduction to tensor networks: Matrix product states and projected entangled pair states},
  author={Or{\'u}s, Rom{\'a}n},
  journal={Annals of physics},
  volume={349},
  pages={117--158},
  year={2014},
  publisher={Elsevier}
}

@article{roggero2021dynamical,
  title={Dynamical phase transitions in models of collective neutrino oscillations},
  author={Roggero, Alessandro},
  journal={Physical Review D},
  volume={104},
  number={12},
  pages={123023},
  year={2021},
  publisher={APS}
}

@article{roggero2021entanglement,
  title={Entanglement and many-body effects in collective neutrino oscillations},
  author={Roggero, Alessandro},
  journal={Physical Review D},
  volume={104},
  number={10},
  pages={103016},
  year={2021},
  publisher={APS}
}

@article{lloyd1996universal,
  title={Universal quantum simulators},
  author={Lloyd, Seth},
  journal={Science},
  volume={273},
  number={5278},
  pages={1073--1078},
  year={1996},
  publisher={American Association for the Advancement of Science}
}

@article{pehlivan2011invariants,
  title={Invariants of collective neutrino oscillations},
  author={Pehlivan, Y and Balantekin, AB and Kajino, Toshitaka and Yoshida, Takashi},
  journal={Physical Review D},
  volume={84},
  number={6},
  pages={065008},
  year={2011},
  publisher={APS}
}

@article{berry2007efficient,
  title={Efficient quantum algorithms for simulating sparse Hamiltonians},
  author={Berry, Dominic W and Ahokas, Graeme and Cleve, Richard and Sanders, Barry C},
  journal={Communications in Mathematical Physics},
  volume={270},
  number={2},
  pages={359--371},
  year={2007},
  publisher={Springer}
}

@article{cirigliano2024neutrino,
  title={Neutrino many-body flavor evolution: The full Hamiltonian},
  author={Cirigliano, Vincenzo and Sen, Srimoyee and Yamauchi, Yukari},
  journal={Physical Review D},
  volume={110},
  number={12},
  pages={123028},
  year={2024},
  publisher={APS}
}

@article{cervia2026interactions,
  title={Interactions of neutrino wave packets},
  author={Cervia, Michael J},
  journal={Physical Review D},
  volume={113},
  number={4},
  pages={043010},
  year={2026},
  publisher={APS}
}

@article{carlson2026neutrino,
  title={Neutrino Flavor Evolution in High Flux Astrophysical Environments},
  author={Carlson, Joseph and Roggero, Alessandro and Neill, Duff},
  journal={arXiv preprint arXiv:2603.12192},
  year={2026}
}

@article{atia2017fast,
  title={Fast-forwarding of Hamiltonians and exponentially precise measurements},
  author={Atia, Yosi and Aharonov, Dorit},
  journal={Nature communications},
  volume={8},
  number={1},
  pages={1572},
  year={2017},
  publisher={Nature Publishing Group UK London}
}

@article{spagnoli2025collective,
  title={Collective neutrino oscillations in three flavors on qubit and qutrit processors},
  author={Spagnoli, Luca and Goss, Noah and Roggero, Alessandro and Rrapaj, Ermal and Cervia, Michael J and Patwardhan, Amol V and Naik, Ravi K and Balantekin, A Baha and Younis, Ed and Santiago, David I and others},
  journal={Physical Review D},
  volume={111},
  number={10},
  pages={103054},
  year={2025},
  publisher={APS}
}

@incollection{patwardhan_ManybodyCollectiveNeutrino_2023,
	title = {Many-body collective neutrino oscillations: recent developments},
	isbn = {978-981-15-8818-1},
	shorttitle = {Many-body collective neutrino oscillations},
	url = {http://arxiv.org/abs/2301.00342},
	urldate = {2023-02-09},
	booktitle = {Handbook of {Nuclear} {Physics}},
	publisher = {Springer, Singapore},
	author = {Patwardhan, Amol V. and Cervia, Michael J. and Rrapaj, Ermal and Siwach, Pooja and Balantekin, A. B.},
	year = {2023},
	note = {10.1007/978-981-15-8818-1\_126-1},
	pages = {1--16},
}

@article{siwach2023collective,
  title={Collective neutrino oscillations on a quantum computer with hybrid quantum-classical algorithm},
  author={Siwach, Pooja and Harrison, Kaytlin and Balantekin, A Baha},
  journal={Physical Review D},
  volume={108},
  number={8},
  pages={083039},
  year={2023},
  publisher={APS}
}

@article{Siwach_2023,
	doi = {10.1103/physrevd.107.023019},
  
	url = {https://doi.org/10.1103%2Fphysrevd.107.023019},
  
	year = 2023,
	month = {jan},
  
	publisher = {American Physical Society ({APS})},
  
	volume = {107},
  
	number = {2},
  
	author = {Pooja Siwach and Anna M. Suliga and A. Baha Balantekin},
  
	title = {Entanglement in three-flavor collective neutrino oscillations},
  
	journal = {Physical Review D}
}

@article{Cervia_2022,
	doi = {10.1103/physrevd.105.123025},
  
	url = {https://doi.org/10.1103%2Fphysrevd.105.123025},
  
	year = 2022,
	month = {jun},
  
	publisher = {American Physical Society ({APS})},
  
	volume = {105},
  
	number = {12},
  
	author = {Michael J. Cervia and Pooja Siwach and Amol V. Patwardhan and A.{\hspace{0.167em}
}B. Balantekin and S.{\hspace{0.167em}}N. Coppersmith and Calvin W. Johnson},
  
	title = {Collective neutrino oscillations with tensor networks using a time-dependent variational principle},
  
	journal = {Physical Review D}
}

@misc{yeteraydeniz2021collective,
      title={Collective Neutrino Oscillations on a Quantum Computer}, 
      author={Kübra Yeter-Aydeniz and Shikha Bangar and George Siopsis and Raphael C. Pooser},
      year={2021},
      eprint={2104.03273},
      archivePrefix={arXiv},
      primaryClass={quant-ph}
}

@article{roggero2022entanglement,
  title={Entanglement and correlations in fast collective neutrino flavor oscillations},
  author={Roggero, Alessandro and Rrapaj, Ermal and Xiong, Zewei},
  journal={Physical Review D},
  volume={106},
  number={4},
  pages={043022},
  year={2022},
  publisher={APS}
}

@article{martin2022classical,
  title={Classical and quantum evolution in a simple coherent neutrino problem},
  author={Martin, Joshua D and Roggero, Alessandro and Duan, Huaiyu and Carlson, J and Cirigliano, Vincenzo},
  journal={Physical Review D},
  volume={105},
  number={8},
  pages={083020},
  year={2022},
  publisher={APS}
}

@article{preskill2018quantum,
  title={Quantum computing in the NISQ era and beyond},
  author={Preskill, John},
  journal={Quantum},
  volume={2},
  pages={79},
  year={2018},
  publisher={Verein zur F{\"o}rderung des Open Access Publizierens in den Quantenwissenschaften}
}

@article{campbell2017roads,
  title={Roads towards fault-tolerant universal quantum computation},
  author={Campbell, Earl T and Terhal, Barbara M and Vuillot, Christophe},
  journal={Nature},
  volume={549},
  number={7671},
  pages={172--179},
  year={2017},
  publisher={Nature Publishing Group UK London}
}

@article{bravyi2005universal,
  title={Universal quantum computation with ideal Clifford gates and noisy ancillas},
  author={Bravyi, Sergey and Kitaev, Alexei},
  journal={Physical Review A—Atomic, Molecular, and Optical Physics},
  volume={71},
  number={2},
  pages={022316},
  year={2005},
  publisher={APS}
}

@article{selinger2015efficient,
  title={Efficient Clifford+ T approximation of single-qubit operators},
  author={Selinger, Peter},
  journal={arXiv preprint arXiv:1212.6253},
  year={2012}
}

@article{low2019hamiltonian,
  title={Hamiltonian simulation by qubitization},
  author={Low, Guang Hao and Chuang, Isaac L},
  journal={Quantum},
  volume={3},
  pages={163},
  year={2019},
  publisher={Verein zur F{\"o}rderung des Open Access Publizierens in den Quantenwissenschaften}
}

@inproceedings{gilyen2019quantum,
  title={Quantum singular value transformation and beyond: exponential improvements for quantum matrix arithmetics},
  author={Gily{\'e}n, Andr{\'a}s and Su, Yuan and Low, Guang Hao and Wiebe, Nathan},
  booktitle={Proceedings of the 51st annual ACM SIGACT symposium on theory of computing},
  pages={193--204},
  year={2019}
}

@article{xiong2022many,
  title={Many-body effects of collective neutrino oscillations},
  author={Xiong, Zewei},
  journal={Physical Review D},
  volume={105},
  number={10},
  pages={103002},
  year={2022},
  publisher={APS}
}

@article{laraib2025many,
  title={Many-Body Simulations of the Fast Flavor Instability},
  author={Laraib, Zoha and Richers, Sherwood},
  journal={arXiv preprint arXiv:2507.02040},
  year={2025}
}

@article{janka_ExplosionMechanismsCoreCollapse_2012,
	title = {Explosion {Mechanisms} of {Core}-{Collapse} {Supernovae}},
	volume = {62},
	url = {https://doi.org/10.1146/annurev-nucl-102711-094901},
	doi = {10.1146/annurev-nucl-102711-094901},
	number = {1},
	urldate = {2021-08-01},
	journal = {Annual Review of Nuclear and Particle Science},
	author = {Janka, Hans-Thomas},
	year = {2012},
	pages = {407--451},
}

@article{mezzacappa_CoreCollapseSupernova_2023,
	title = {Core collapse supernova gravitational wave emission for progenitors of 9.6, 15, and 25 {Solar} {Masses}},
	volume = {107},
	url = {https://link.aps.org/doi/10.1103/PhysRevD.107.043008},
	doi = {10.1103/PhysRevD.107.043008},
	number = {4},
	urldate = {2023-06-25},
	journal = {Physical Review D},
	author = {Mezzacappa, Anthony and Marronetti, Pedro and Landfield, Ryan E. and Lentz, Eric J. and Murphy, R. Daniel and Raphael Hix, W. and Harris, J. Austin and Bruenn, Stephen W. and Blondin, John M. and Bronson Messer, O.E. and Casanova, Jordi and Kronzer, Luke L.},
	month = feb,
	year = {2023},
	note = {Publisher: American Physical Society},
	pages = {043008},
}

@article{foucart2023neutrino,
  title={Neutrino transport in general relativistic neutron star merger simulations},
  author={Foucart, Francois},
  journal={Living Reviews in Computational Astrophysics},
  volume={9},
  number={1},
  pages={1},
  year={2023},
  publisher={Springer}
}

@article{burrows2021core,
  title={Core-collapse supernova explosion theory},
  author={Burrows, Adam and Vartanyan, David},
  journal={Nature},
  volume={589},
  number={7840},
  pages={29--39},
  year={2021},
  publisher={Nature Publishing Group UK London}
}

@article{Qian1993,
  author = {Qian, Yong-Zhong and Woosley, S. E.},
  title = {Nucleosynthesis in Neutrino-driven Winds. I. The Physical Conditions},
  journal = {Astrophys. J.},
  year = {1996},
  volume = {471},
  pages = {331},
}

@incollection{wang2022neutrinos,
  title={Neutrinos and Heavy Element Nucleosynthesis},
  author={Wang, Xilu and Surman, Rebecca},
  booktitle={Handbook of Nuclear Physics},
  pages={1--19},
  year={2022},
  publisher={Springer}
}

@article{fischer2024neutrinos,
  title={Neutrinos and nucleosynthesis of elements},
  author={Fischer, Tobias and Guo, Gang and Langanke, Karlheinz and Mart{\'\i}nez-Pinedo, Gabriel and Qian, Yong-Zhong and Wu, Meng-Ru},
  journal={Progress in Particle and Nuclear Physics},
  pages={104107},
  year={2024},
  publisher={Elsevier}
}

@incollection{richers_FastFlavorTransformations_2022,
	title = {Fast {Flavor} {Transformations}},
	copyright = {All rights reserved},
	isbn = {978-981-15-8818-1},
	booktitle = {Handbook of {Nuclear} {Physics}},
	publisher = {Springer, Singapore},
	author = {Richers, Sherwood and Sen, Manibrata},
	month = jul,
	year = {2022},
}

@article{li2021neutrino,
  title={Neutrino fast flavor conversions in neutron-star postmerger accretion disks},
  author={Li, Xinyu and Siegel, Daniel M},
  journal={Physical Review Letters},
  volume={126},
  number={25},
  pages={251101},
  year={2021},
  publisher={APS}
}

@article{just2022fast,
  title={Fast neutrino conversion in hydrodynamic simulations of neutrino-cooled accretion disks},
  author={Just, Oliver and Abbar, Sajad and Wu, Meng-Ru and Tamborra, Irene and Janka, Hans-Thomas and Capozzi, Francesco},
  journal={Physical Review D},
  volume={105},
  number={8},
  pages={083024},
  year={2022},
  publisher={APS}
}

@article{chakraborty2016self2,
  title={Self-induced neutrino flavor conversion without flavor mixing},
  author={Chakraborty, Sovan and Hansen, Rasmus Sloth and Izaguirre, Ignacio and Raffelt, Georg G},
  journal={Journal of Cosmology and Astroparticle Physics},
  volume={2016},
  number={03},
  pages={042},
  year={2016},
  publisher={IOP Publishing}
}

@article{fernandez_FastFlavorInstability_2022a,
	title = {Fast flavor instability in hypermassive neutron star disk outflows},
	volume = {106},
	copyright = {All rights reserved},
	url = {https://link.aps.org/doi/10.1103/PhysRevD.106.103003},
	doi = {10.1103/PhysRevD.106.103003},
	number = {10},
	urldate = {2022-11-11},
	journal = {Physical Review D},
	author = {Fernández, Rodrigo and Richers, Sherwood and Mulyk, Nicole and Fahlman, Steven},
	month = nov,
	year = {2022},
	note = {Publisher: American Physical Society},
	pages = {103003},
}

@article{ehring2023fast,
  title={Fast neutrino flavor conversions can help and hinder neutrino-driven explosions},
  author={Ehring, Jakob and Abbar, Sajad and Janka, Hans-Thomas and Raffelt, Georg and Tamborra, Irene},
  journal={Physical Review Letters},
  volume={131},
  number={6},
  pages={061401},
  year={2023},
  publisher={APS}
}

@article{wang2025effect,
  title={The Effect of the Fast-flavor Instability on Core-collapse Supernova Models},
  author={Wang, Tianshu and Burrows, Adam},
  journal={The Astrophysical Journal},
  volume={986},
  number={2},
  pages={153},
  year={2025},
  publisher={IOP Publishing}
}

@article{dasgupta2017fast,
  title={Fast neutrino flavor conversions near the supernova core with realistic flavor-dependent angular distributions},
  author={Dasgupta, Basudeb and Mirizzi, Alessandro and Sen, Manibrata},
  journal={Journal of Cosmology and Astroparticle Physics},
  volume={2017},
  number={02},
  pages={019},
  year={2017},
  publisher={IOP Publishing}
}

@article{banerjee2011linearized,
  title={Linearized flavor-stability analysis of dense neutrino streams},
  author={Banerjee, Arka and Dighe, Amol and Raffelt, Georg},
  journal={Physical Review D—Particles, Fields, Gravitation, and Cosmology},
  volume={84},
  number={5},
  pages={053013},
  year={2011},
  publisher={APS}
}

@article{illa2023multi,
  title={Multi-neutrino entanglement and correlations in dense neutrino systems},
  author={Illa, Marc and Savage, Martin J},
  journal={Physical Review Letters},
  volume={130},
  number={22},
  pages={221003},
  year={2023},
  publisher={APS}
}

@article{Shalgar2021breaking,
  author = {S. Shalgar and I. Tamborra},
  title = {Breaking the symmetries in self-interacting neutrino systems: Fast conversions and directional crossings},
  journal = {Phys. Rev. D},
  volume = {103},
  number = {6},
  pages = {063002},
  year = {2021},
}

@article{fiorillo2024fast,
  title={Fast flavor conversions at the edge of instability in a two-beam model},
  author={Fiorillo, Damiano FG and Raffelt, Georg G},
  journal={Physical Review Letters},
  volume={133},
  number={22},
  pages={221004},
  year={2024},
  publisher={APS}
}

@article{fiorillo2024inhomogeneous,
  title={Inhomogeneous kinetic equation for mixed neutrinos: tracing the missing energy},
  author={Fiorillo, Damiano FG and Raffelt, Georg G and Sigl, G{\"u}nter},
  journal={Physical Review Letters},
  volume={133},
  number={2},
  pages={021002},
  year={2024},
  publisher={APS}
}

@article{wu2017imprints,
  title={Imprints of neutrino-pair flavor conversions on nucleosynthesis in ejecta from neutron-star merger remnants},
  author={Wu, Meng-Ru and Tamborra, Irene and Just, Oliver and Janka, Hans-Thomas},
  journal={Physical Review D},
  volume={96},
  number={12},
  pages={123015},
  year={2017},
  publisher={APS}
}

@article{johns2025neutrino,
  title={Neutrino Oscillations in Core-Collapse Supernovae and Neutron Star Mergers},
  author={Johns, Lucas and Richers, Sherwood and Wu, Meng-Ru},
  journal={Annual Review of Nuclear and Particle Science},
  volume={75},
  year={2025},
  publisher={Annual Reviews}
}

@article{volpe2024neutrinos,
  title={Neutrinos from dense environments: Flavor mechanisms, theoretical approaches, observations, and new directions},
  author={Volpe, M Cristina},
  journal={Reviews of Modern Physics},
  volume={96},
  number={2},
  pages={025004},
  year={2024},
  publisher={APS}
}

@article{balantekin2024collective,
  title={Collective neutrino oscillations and heavy-element nucleosynthesis in supernovae: Exploring potential effects of many-body neutrino correlations},
  author={Balantekin, A Baha and Cervia, Michael J and Patwardhan, Amol V and Surman, Rebecca and Wang, Xilu},
  journal={The Astrophysical Journal},
  volume={967},
  number={2},
  pages={146},
  year={2024},
  publisher={The American Astronomical Society}
}

@article{low2017optimal,
  title = {Optimal Hamiltonian Simulation by Quantum Signal Processing},
  author = {Low, Guang Hao and Chuang, Isaac L.},
  journal = {Phys. Rev. Lett.},
  volume = {118},
  issue = {1},
  pages = {010501},
  numpages = {5},
  year = {2017},
  month = {Jan},
  publisher = {American Physical Society},
  doi = {10.1103/PhysRevLett.118.010501},
  url = {https://link.aps.org/doi/10.1103/PhysRevLett.118.010501}
}

@article{bauer2023hep,
  title = {Quantum Simulation for High-Energy Physics},
  author = {Bauer, Christian W. and Davoudi, Zohreh and Balantekin, A. Baha and Bhattacharya, Tanmoy and Carena, Marcela and de Jong, Wibe A. and Draper, Patrick and El-Khadra, Aida and Gemelke, Nate and Hanada, Masanori and Kharzeev, Dmitri and Lamm, Henry and Li, Ying-Ying and Liu, Junyu and Lukin, Mikhail and Meurice, Yannick and Monroe, Christopher and Nachman, Benjamin and Pagano, Guido and Preskill, John and Rinaldi, Enrico and Roggero, Alessandro and Santiago, David I. and Savage, Martin J. and Siddiqi, Irfan and Siopsis, George and Van Zanten, David and Wiebe, Nathan and Yamauchi, Yukari and Yeter-Aydeniz, K\"ubra and Zorzetti, Silvia},
  journal = {PRX Quantum},
  volume = {4},
  issue = {2},
  pages = {027001},
  numpages = {70},
  year = {2023},
  month = {May},
  publisher = {American Physical Society},
  doi = {10.1103/PRXQuantum.4.027001},
  url = {https://link.aps.org/doi/10.1103/PRXQuantum.4.027001}
}

@article{bosse2025efficient,
  title={Efficient and practical Hamiltonian simulation from time-dependent product formulas},
  author={Bosse, Jan Lukas and Childs, Andrew M and Derby, Charles and Gambetta, Filippo Maria and Montanaro, Ashley and Santos, Raul A},
  journal={Nature Communications},
  volume={16},
  number={1},
  pages={2673},
  year={2025},
  publisher={Nature Publishing Group UK London}
}

@article{clinton2021hamiltonian,
  title={Hamiltonian simulation algorithms for near-term quantum hardware},
  author={Clinton, Laura and Bausch, Johannes and Cubitt, Toby},
  journal={Nature communications},
  volume={12},
  number={1},
  pages={4989},
  year={2021},
  publisher={Nature Publishing Group UK London}
}

@incollection{barrett2013codesign,
  author    = {Richard F. Barrett and Shekhar Borkar and Sudip S. Dosanjh and Simon D. Hammond and Michael A. Heroux and X. Sharon Hu and Justin Luitjens and Steven G. Parker and John Shalf and Li Tang},
  title     = {On the role of co-design in high performance computing},
  booktitle = {Transition of HPC Towards Exascale Computing},
  editor    = {Jack J. Dongarra and Ian Foster and Lucio Grandinetti},
  publisher = {SAGE Publications, Limited},
  year      = {2013},
  pages     = {141--155},
  edition   = {1}
}

@INPROCEEDINGS{shankar2021codesign,
  author={Shankar, Sadasivan Sadas},
  booktitle={2021 IEEE High Performance Extreme Computing Conference (HPEC)}, 
  title={Lessons from Nature for Computing: Looking beyond Moore’s Law with Special Purpose Computing and Co-design}, 
  year={2021},
  volume={},
  number={},
  pages={1-8},
  doi={10.1109/HPEC49654.2021.9622865}}

@article{gibbs2025deep,
  title={Deep circuit compression for quantum dynamics via tensor networks},
  author={Gibbs, Joe and Cincio, Lukasz},
  journal={Quantum},
  volume={9},
  pages={1789},
  year={2025},
  publisher={Verein zur F{\"o}rderung des Open Access Publizierens in den Quantenwissenschaften}
}

@article{kemper2024denoising,
  title = {Denoising and Extension of Response Functions in the Time Domain},
  author = {Kemper, Alexander F. and Yang, Chao and Gull, Emanuel},
  journal = {Phys. Rev. Lett.},
  volume = {132},
  issue = {16},
  pages = {160403},
  numpages = {7},
  year = {2024},
  month = {Apr},
  publisher = {American Physical Society},
  doi = {10.1103/PhysRevLett.132.160403},
  url = {https://link.aps.org/doi/10.1103/PhysRevLett.132.160403}
}

@article{hall2021simulation,
  title={Simulation of collective neutrino oscillations on a quantum computer},
  author={Hall, Benjamin and Roggero, Alessandro and Baroni, Alessandro and Carlson, Joseph},
  journal={Physical review D},
  volume={104},
  number={6},
  pages={063009},
  year={2021},
  publisher={APS}
}

@article{laraib2025two,
  title={Two-beam multiparticle many-body simulations of inhomogeneous fast flavor instabilities},
  author={Laraib, Zoha and Richers, Sherwood},
  journal={Physical Review D},
  volume={112},
  number={12},
  pages={123045},
  year={2025},
  publisher={APS}
}

@article{Mooney_2021,
   title={Cost-optimal single-qubit gate synthesis in the Clifford hierarchy},
   volume={5},
   ISSN={2521-327X},
   url={http://dx.doi.org/10.22331/q-2021-02-15-396},
   DOI={10.22331/q-2021-02-15-396},
   journal={Quantum},
   publisher={Verein zur Forderung des Open Access Publizierens in den Quantenwissenschaften},
   author={Mooney, Gary J. and Hill, Charles D. and Hollenberg, Lloyd C. L.},
   year={2021},
   month=Feb, pages={396} }

@article{chernyshev2025quantum,
  title={Quantum magic and computational complexity in the neutrino sector},
  author={Chernyshev, Ivan and Robin, Caroline EP and Savage, Martin J},
  journal={Physical Review Research},
  volume={7},
  number={2},
  pages={023228},
  year={2025},
  publisher={APS}
}

@article{hirano2024magicpool,
  title={Magicpool: Dealing with magic state distillation failures on large-scale fault-tolerant quantum computer},
  author={Hirano, Yutaka and Suzuki, Yasunari and Fujii, Keisuke},
  journal={arXiv preprint arXiv:2407.07394},
  year={2024}
}

@article{acharya2024quantum,
  title={Quantum error correction below the surface code threshold},
  author={Acharya, Rajeev and Aghababaie-Beni, Laleh and Aleiner, Igor and Andersen, Trond I and Ansmann, Markus and Arute, Frank and Arya, Kunal and Asfaw, Abraham and Astrakhantsev, Nikita and Atalaya, Juan and others},
  journal={arXiv preprint arXiv:2408.13687},
  year={2024}
}

@article{jacinto2026exploring,
  title={Exploring the landscape of compact magic-state distillation factories},
  author={Jacinto, Hugo and Valcarce, Xavier and Barizien, Victor and Gouzien, {\'E}lie and Sangouard, Nicolas},
  journal={arXiv preprint arXiv:2606.07734},
  year={2026}
}

@article{wills2025constant,
  title={Constant-overhead magic state distillation},
  author={Wills, Adam and Hsieh, Min-Hsiu and Yamasaki, Hayata},
  journal={Nature Physics},
  volume={21},
  number={11},
  pages={1842--1846},
  year={2025},
  publisher={Nature Publishing Group UK London}
}

@article{o2017quantum,
  title={Quantum computation with realistic magic-state factories},
  author={O'Gorman, Joe and Campbell, Earl T},
  journal={Physical Review A},
  volume={95},
  number={3},
  pages={032338},
  year={2017},
  publisher={APS}
}

@article{koenig2024inverted,
  title={Inverted-circuit zero-noise extrapolation for quantum-gate error mitigation},
  author={Koenig, Kathrin F and Reinecke, Finn and Hahn, Walter and Wellens, Thomas},
  journal={Physical Review A},
  volume={110},
  number={4},
  pages={042625},
  year={2024},
  publisher={APS}
}

@article{froustey2026collective,
  title={Collective neutrino oscillations: Many-body non-forward effects and non-classicality},
  author={Froustey, Julien and Rrapaj, Ermal and Liu, Yuhao and Li, Gushu and Iancu, Costin and Cirigliano, Vincenzo},
  journal={arXiv preprint arXiv:2606.12404},
  year={2026}
}

@article{heimsoth2026three,
  title={Three-flavor supernova neutrino simulation using a hybrid quantum-classical algorithm with qutrits},
  author={Heimsoth, Daniel J and Siwach, Pooja and Balantekin, A Baha},
  journal={Physical Review D},
  volume={114},
  number={4},
  pages={043055},
  year={2026},
  publisher={APS}
}

@article{vlasenko2014neutrino,
  title={Neutrino quantum kinetics},
  author={Vlasenko, Alexey and Fuller, George M and Cirigliano, Vincenzo},
  journal={Physical Review D},
  volume={89},
  number={10},
  pages={105004},
  year={2014},
  publisher={APS}
}

\end{document}